\documentclass[sigconf]{acmart}

\usepackage{wasysym}
\usepackage{geometry}
\usepackage{graphicx}
\usepackage{subfig}
\usepackage{caption}
\usepackage{hyperref}
\usepackage{amsmath}
\usepackage{multirow}
\usepackage{xspace}
\usepackage{threeparttable}
\usepackage{balance}
\usepackage{enumitem}
\let\oldhat\hat
    
\renewcommand{\hat}[1]{\oldhat{\mathbf{#1}}}

\newcommand{\ie}{\emph{i.e., }}
\newcommand{\eg}{\emph{e.g., }}

\newcommand{\wrt}{\emph{w.r.t. }}

\newcommand{\aka}{\emph{aka. }}

\AtBeginDocument{%
  \providecommand\BibTeX{{%
    \normalfont B\kern-0.5em{\scshape i\kern-0.25em b}\kern-0.8em\TeX}}}

\copyrightyear{2022}
\acmYear{2022}
\setcopyright{acmcopyright}
\acmConference[CIKM '22] {Proceedings of the 31st ACM International Conference on Information and Knowledge Management}{October 17--21, 2022}{Atlanta, GA, USA.}
\acmBooktitle{Proceedings of the 31st ACM International Conference on Information and Knowledge Management (CIKM '22), October 17--21, 2022, Atlanta, GA, USA}
\acmPrice{15.00}
\acmISBN{978-1-4503-9236-5/22/10}
\acmDOI{10.1145/3511808.3557358}

\begin{document}

\title{Improving Knowledge-aware Recommendation with Multi-level Interactive Contrastive Learning}


\author{Ding Zou}
\affiliation{%
	\institution{CCIIP Laboratory, Huazhong University of Science and Technology}
	\institution{Joint Laboratory of HUST and Pingan Property \& Casualty Research (HPL)}
	\country{China}
}
\email{m202173662@hust.edu.cn}

\author{Wei Wei}
\authornote{Corresponding author.}
\affiliation{%
	\institution{CCIIP Laboratory, Huazhong University of Science and Technology}
	\institution{Joint Laboratory of HUST and Pingan Property \& Casualty Research (HPL)}
	\country{China}
}
\email{weiw@hust.edu.cn}

\author{Ziyang Wang}
\affiliation{%
	\institution{CCIIP Laboratory, Huazhong University of Science and Technology}
	\institution{Joint Laboratory of HUST and Pingan Property \& Casualty Research (HPL)}
	\country{China}
}
\email{ziyang1997@hust.edu.cn}

\author{Xian-Ling Mao}
\affiliation{%
	\institution{Beijing Institute of Technology}
	\country{China}
}
\email{maoxl@bit.edu.cn}

\author{Feida Zhu}
\affiliation{%
	\institution{Singapore Management University}
	\country{Singapore}
}
\email{fdzhu@smu.edu.sg}

\author{Fang Rui}
\affiliation{%
	\institution{Ping An Property \& Casualty Insurance company of China, Ltd}
	\country{China}
}
\email{fangrui051@pingan.com.cn}

\author{Dangyang Chen}
\affiliation{%
	\institution{Ping An Property \& Casualty Insurance company of China, Ltd}
	\country{China}
}
\email{chendangyang273@pingan.com.cn}

\begin{abstract}
Incorporating Knowledge Graphs (KG) into recommeder system as side information has attracted considerable attention. Recently, the technical trend of Knowledge-aware Recommendation (KGR) is to develop end-to-end models based on graph neural networks (GNNs). 
However, the extremely sparse user-item interactions significantly degrade the performance of the GNN-based models, from the following aspects:
1) the sparse interaction, itself, means inadequate supervision signals and limits the supervised GNN-based models;
2) the combination of sparse interactions (CF part) and redundant KG facts (KG part) further results in an unbalanced information utilization.
Besides, the GNN paradigm aggregates local neighbors for node representation learning, while ignoring the non-local KG facts and making the knowledge extraction insufficient.
Inspired by the recent success of contrastive learning in mining supervised signals from data itself, in this paper, we focus on exploring contrastive learning in KGR and propose a novel multi-level interactive contrastive learning mechanism, to alleviate the aforementioned challenges. 
Different from traditional contrastive learning methods which contrast nodes of two generated graph views, interactive contrastive mechanism conducts layer-wise self-supervised learning by contrasting layers of different parts within graphs, which is also an "interaction" action. 
Specifically, we first construct local and non-local graphs for user/item in KG, exploring more KG facts for KGR. Then an intra-graph level interactive contrastive learning is performed within each local/non-local graph, which contrasts layers of the CF and KG parts, for more consistent information leveraging. Besides, an inter-graph level interactive contrastive learning is performed between the local and non-local graphs, for sufficiently and coherently extracting non-local KG signals.
Extensive experiments conducted on three benchmark datasets show the superior performance of our proposed method over the state-of-the-arts.
The implementations are available at: https://github.com/CCIIPLab/KGIC.
\end{abstract}


\maketitle

\section{Introduction}
\begin{figure}[t]
  \centering
  \includegraphics[width=\linewidth]{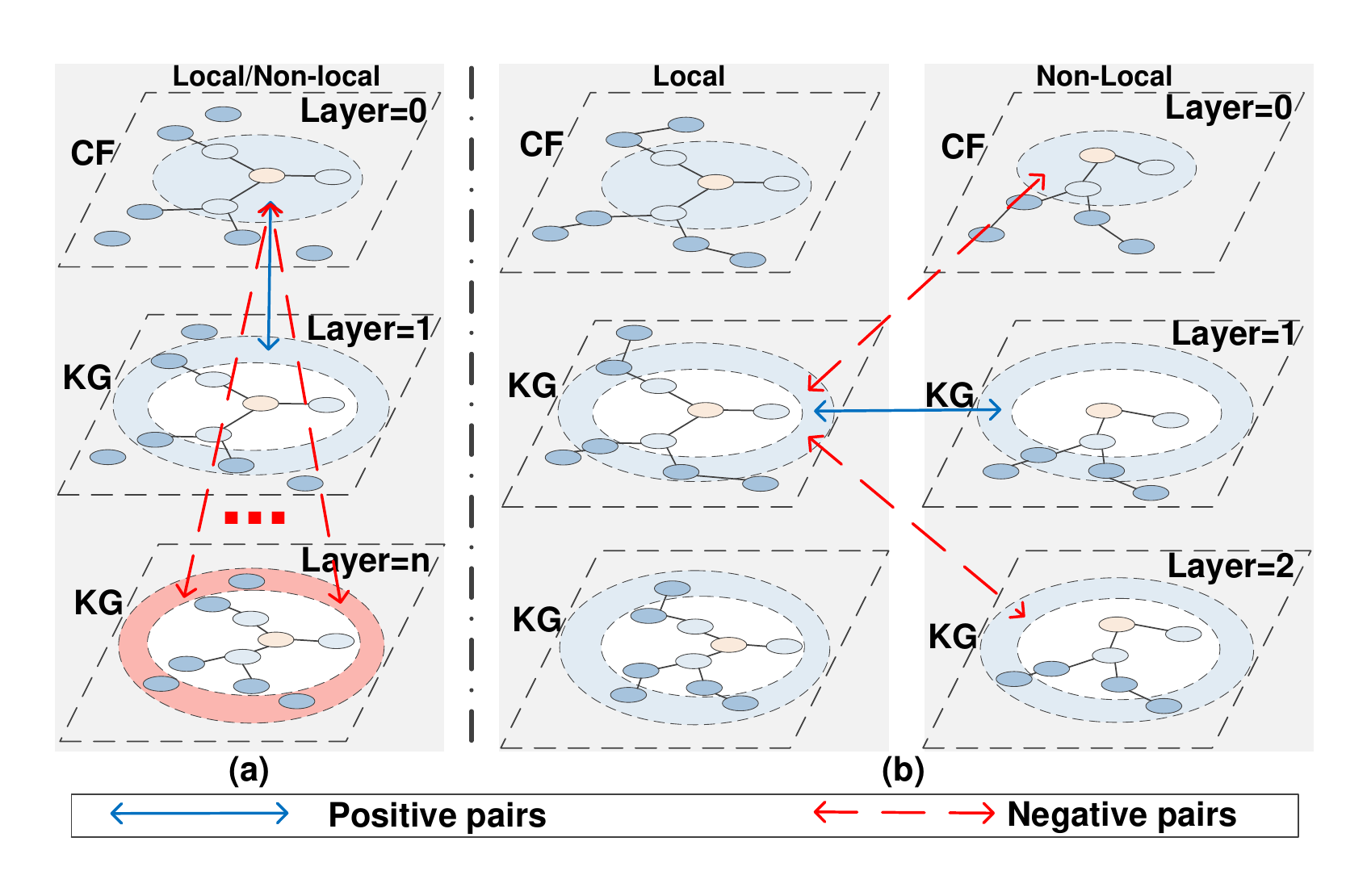}
  \caption{A toy example of multi-level contrastive mechanism. (a) Intra-graph level interactive contrastive mechanism; (b) Inter-graph level interactive contrastive mechanism (Only taking Layer=1 as the anchor here for clarity).}
  \label{fig:toy}
\end{figure}

Recommender systems (RS) aim to alleviate the information explosion, proposing to recommend a small set of items to meet users' personalized interests. As an effective solution, collaborative filtering \cite{he2017neural,lian2018xdeepfm, PengSM22, zhao22AAAI, wang2020global} presumes that behaviorally similar users have a similar preference on items and achieves great success. However, they severely suffer from the cold-start problem, due to treating each interaction as an independent instance while neglecting their relations. A commonly-adopted solution is to integrate side information with collaborative filtering models, such as a knowledge graph (KG) \cite{pan2021context, Zou2022multilevel}, which contains rich facts and connections about items, to learn high-quality user and item representations for recommendation (\aka knowledge-aware recommendation, KGR).

Indeed, there already exists much research effort \cite{wang2018dkn, zhang2016collaborative, zhang2018learning} for KGR, targeting to \textbf{sufficiently and coherently} leverage the graph information of CF (\ie user-item interactions \cite{liu2014exploiting}) and KG (\ie item-entity affiliations). 
Earlier studies \cite{zhang2016collaborative, huang2018improving,wang2018dkn} explore CF and KG independently, which exploits KG to enhance item representation in CF. They employ \emph{knowledge graph embedding} (KGE) models (\eg TransE \cite{bordes2013translating}, TransH \cite{wang2014knowledge}) to pre-train entity embeddings and treat them as prior information. However, these methods fall short in distilling sufficient knowledge signals from KG, since they treat each item-entity relation independently.
Hence, some follow-on studies \cite{hu2018leveraging, shi2018heterogeneous, wang2019explainable} focus on extracting more sufficient KG signals to enrich CF, via capturing the long-range KG connectivity, such as selecting prominent paths over KG \cite{sun2018recurrent} \emph{or} representing the interactions with multi-hop paths from users to items \cite{hu2018leveraging, wang2019explainable}. Nevertheless, most of them heavily rely on manually designed meta-paths, thus are hard to optimize in reality.
More recent works \cite{sha2019attentive, wang2019knowledge, wang2019kgat, wang2021learning} further unify CF and KG as a heterogeneous graph, with an informative aggregation paradigm (\ie Graph Neural Networks, GNNs) to perform graph representation learning. Due to the powerful capability of GNNs in effectively generating local permutation-invariant aggregation on the neighbors of a node, these methods achieve promising performance and become the mainstream tendency in KGR.

Despite effectiveness, existing GNN-based methods still fall short in achieving the above core goal in the following three aspects:
\begin{itemize}[leftmargin=*]
    \item \textbf{Sparse Supervised Signals.}
    Since established in a supervised manner, existing GNN-based methods rely on the observed user-item interactions as supervision signals to perform graph representation learning on the unified heterogeneous graph. However, the user-item interactions are actually extremely sparse in real scenarios \cite{bayer2017generic, wu2021self}, which makes it insufficient to achieve satisfactory performance and even results in terrible side effects, \eg degeneration problem \cite{gao2019representation} (\ie degenerating node embeddings distribution into a narrow cone, even leading to the indiscrimination of generated node representations).
    \item \textbf{Unbalanced Information Utilization.}
    When the sparse user-item interactions meet redundant knowledge facts, there occurs an unbalanced heterogeneous structure in KGR, which results in an unbalanced information utilization problem. A commonly approved fact is that, it's the CF signals that determine the user's preferences, since they are composed of user's historical interactions. With the unbalanced information utilization, the noisy KG information, however, gets emphasized more in final user/item modeling, which makes the crucial CF signals less stressed and further results in suboptimal representation learning.   
    \item \textbf{Insufficient Knowledge Extraction.} 
    Although there are redundant KG facts, extracting informative and helpful knowledge is actually far from sufficient in previous GNN-based models, due to the local aggregation feature of GNN.
    The GNN-based methods usually learn item representations by aggregating neighboring entities on their local KG structures (\ie neighboring areas of the item itself), which ignores the non-local KG facts (\ie neighboring areas of similar items) and thus results in inadequate knowledge extraction. Nevertheless, simply aggregating the non-local KG facts may introduce more irrelevant noise, further resulting in an unexpected performance decrease. 
\end{itemize}

Inspired by recent success in contrastive learning, one of the classical Self-Supervised Learning (SSL) methods, we naturally propose to leverage the superiority of SSL to alleviate the first problem of Sparse Supervised Signals. 
A straightforward idea is that we could augment (or corrupt) the input user-item-entity graph as a graph view, and contrast the nodes in it with the original one, following the traditional contrastive paradigm analogous to \cite{chen2020simple, he2020momentum, lan2019albert}. 
However, such a paradigm performs contrastive learning in a relatively independent manner, which only contrasts the same part (CF or KG part) of different graph views, hence ignores the inner interactions of different parts within a graph. That is, the improvements of CF and KG representation learning are mutually isolated, and the impact of CF part in final user/item modeling is still limited. Hence with the traditional contrastive mechanism, the above second and third problems are still far from resolution or alleviation.
As a result, it is a necessity to endow the contrastive learning paradigm with the capability of effective \emph{information interaction} between CF and KG parts, so as to coherently exploit information of each part without reliance on extra explicit labels. 
This motivates us to design an interactive graph contrastive mechanism tailored for KGR task, which is required to conclude the following aspects so as to address the above limitations: 
1) contrasting the CF and KG parts to balance their impact on representation learning, as shown in Figure \ref{fig:toy} (a);
2) contrasting the local and non-local graphs in KG to extract informative non-local KG facts, as shown in Figure \ref{fig:toy} (b).

In this paper, we develop a novel model, \underline{K}nowled\underline{g}e-aware Recommender System with Multi-level \underline{I}nteractive \underline{C}ontrastive Learning (KGIC), to solve the aforementioned limitations and challenges. Conceptually, KGIC focuses on exploring a proper in-graph contrastive learning mechanism for KGR, aiming to unify the two crucial but relatively independent parts (\ie CF and KG) in a self-supervised manner. In particular, we first perform the interactive graph contrastive learning in \textbf{intra-graph level}, which contrasts the CF signals with the KG information in each local/non-local graph of user/item, aiming to increase the information consistency between CF and KG parts. 
Then an \textbf{inter-graph level} interactive graph contrastive learning is conducted, which contrasts the local KG facts with non-local ones, for integrating more KG signals and denoising the non-local KG information.
Empirically, KGIC outperforms the state-of-the-art models on three benchmark datasets.

\textbf{Our contributions} of this work can be summarized as follows:
\begin{itemize}[leftmargin=*]
    \item \textbf{General Aspects:} We emphasize the importance of incorporating self-supervised learning to unify CF and KG information for knowledge-aware recommendation, which takes layer self-discrimination as a self-supervised task to offer auxiliary signals for coherent graph representation learning.
    \item \textbf{Novel Methodologies:}
    We propose a novel model KGIC, which designs a multi-level interactive contrastive mechanism for knowledge-aware recommendation. KGIC combines multi-order CF with KG to construct local and non-local graphs for fully exploring external knowledge. KGIC then performs intra-graph level and inter-graph level interactive contrastive learning among local and non-local graphs, for a sufficient and coherent information utilization in CF and KG.
    \item \textbf{Multifaceted Experiments:} We conduct extensive experiments on three benchmark datasets. The results demonstrate the advantages of our KGIC in better representation learning, which shows the effectiveness of our multi-level interactive contrastive learning for KGR.
\end{itemize}

\section{Related Work}

\subsection{Knowledge-aware Recommendation}

\subsubsection{Embedding-based methods.} \label{embedding-based}
 Embedding-based methods \cite{cao2019unifying, wang2018dkn, zhang2016collaborative, huang2018improving, zhang2018learning, wang2018shine, wang2019multi} pre-train the KG entity embeddings with knowledge graph embeddings methods (KGE) \cite{wang2014knowledge, bordes2013translating, lin2015learning}, for enriching item representations. CKE \cite{zhang2016collaborative} combines CF module with structural, textual, and visual knowledge embeddings of items in a unified Bayesian framework. KTUP \cite{cao2019unifying} jointly model user preference and perform KG completion with the TransH \cite{wang2014knowledge} method on user-item interactions and KG triplets. RippleNet \cite{wang2018ripplenet} propagates users’ historical clicked items along links in KG, exploring more KG facts to reveal the users' potential interests. Embedding-based methods show high flexibility in utilizing KG, but the utilized KGE focus more on modeling rigorous semantic relatedness, which is more suitable for link prediction rather than recommendation.

\subsubsection{Path-based methods.} \label{path-based}
Path-based methods \cite{yu2014personalized, hu2018leveraging, shi2018heterogeneous, wang2019explainable, yu2013collaborative, zhao2017meta} explore various patterns of connections among items in KG to provide additional guidance for the recommendation. PER \cite{yu2014personalized} and meta-graph based recommendation \cite{hu2018leveraging} extract the meta-path/meta-graph latent features and exploit the connectivity between users and items along different types of relation paths/graphs. KPRN \cite{wang2019explainable} further automatically extracts paths between users and items, modeling these paths with RNNs. 
Path-based methods mostly involve the design of meta-paths for generating meaningful connectivity patterns, which requires specific domain knowledge and labor-intensive human efforts for accurate path construction.

\subsubsection{GNN-based methods.}
GNN-based methods \cite{yu2014personalized, hu2018leveraging, wang2019explainable, yu2013collaborative, zhao2017meta} are founded on the information aggregation mechanism of graph neural networks (GNNs) \cite{kipf2016semi, hamilton2017inductive, ying2018graph}. 
Typically it integrates multi-hop neighbors into node representations to capture node feature and graph structure, modeling long-range connectivity. KGCN \cite{wang2019knowledge} and KGNN-LS \cite{wang2019knowledge-aware} firstly utilize GNN on KG to obtain item embeddings by aggregating items’ neighborhood information iteratively. Later, KGAT \cite{wang2019kgat} combines the user-item graph with the KG as a unified heterogeneous graph, then utilizes GNN to recursively perform aggregation on it. But CKAN \cite{wang2020ckan} separately propagates collaborative signals and knowledge signals on the user-item graph and KG to stress the importance of CF signals. More recently, KGIN \cite{wang2021learning} models user-item interactions at an intent level, which reveals user intents behind the KG interactions and performs GNN on the user-intent-item-entity graph. And CG-KGR \cite{chen2022attentive} exploits the pre-trained collaborative signals to guide the aggregation on KG, for sufficient knowledge extraction.
However, all these approaches follow the supervised learning paradigm, hence suffering from the original sparse interactions. Moreover, most of them have an unbalanced utilization of user-item-entity graph, overstressing the importance of KG and ignoring the crucial effect of CF.
Besides, insufficient knowledge extraction also exists for the local aggregation feature of GNN.
In contrast, our work explores self-supervised learning in KGR, to overcome the limitations brought by the spare interactions and achieve sufficient knowledge extraction. 

\subsection{Contrastive Learning}
Contrastive learning methods \cite{velickovic2019deep, wu2021self, wang2021self} learn node representations by contrasting positive pairs against negative pairs.
DGI \cite{velickovic2019deep} first adopts Infomax \cite{linsker1988self} in graph representation learning, and contrasts the local node embeddings with global graph embeddings. Then GMI \cite{peng2020graph} proposes to contrast the center node with its local neighbors on both node features and topological structure. Similarly, MVGRL \cite{hassani2020contrastive} learns node-level and graph-level node representations from first-order neighbors and a graph diffusion, and contrasts encoded embeddings between two graph views. More recently, HeCo \cite{wang2021self} proposes to learn node representations from network schema view and meta-path view, and performs contrastive learning between them. As for contrastive learning in the traditional CF-based recommendation domain, SGL \cite{wu2021self} generates two graph views by corrupting the user-item interactions, and performs contrastive learning between them. And NCL \cite{lin2022ncl} captures potential node (structural and semantic) relatedness into contrastive learning for a neighbor-enhanced contrasting.
However, little effort has been made to investigate the great potential of contrastive learning on knowledge-aware recommendation.
\section{Problem Formulation}

In this section, we first formulate structural data of CF and KG parts, \ie user-item interactions and knowledge graph, then present the problem statement of knowledge-aware recommendation.

\noindent\textbf{Interaction Data}. 
In a typical recommendation scenario, let $\mathcal{U}=\left\{u_{1}, u_{2}, \ldots, u_{M}\right\}$ be a set of $M$ users and $\mathcal{V}=\left\{v_{1}, v_{2}, \ldots, v_{N}\right\}$ a set of $N$ items.
Let $\mathbf{Y} \in \mathbf{R}^{M \times N}$ be the user-item interaction matrix, where $y_{u v}=1$ indicates that user $u$ engaged with item $v$, such as behaviors like clicking or purchasing; otherwise $y_{u v}=0$.

\noindent\textbf{Knowledge Graph}. 
A KG stores luxuriant real-world facts associated with items, \eg item attributes, or external commonsense knowledge,  in the form of a heterogeneous graph \cite{shi2018heterogeneous}. 
Let $\mathcal{G}=\{(h, r, t) \mid h, t \in \mathcal{E}, r \in \mathcal{R}\}$ be the knowledge graph, where $h$, $r$, $t$ are on behalf of head, relation, tail of a knowledge triple correspondingly; $\mathcal{E}$ and $\mathcal{R}$ refer to the sets of entities and relations in $\mathcal{G}$. 
In many recommendation scenarios, an item $v \in \mathcal{V}$ corresponds to one entity $e \in \mathcal{E}$. So we establish a set of item-entity alignments $\mathcal{A} =\{(v, e)|v \in \mathcal{V}, e \in \mathcal{E}\}$, where $\left(v, e\right)$ indicates that item $v$ can be aligned with an entity $e$ in the KG.
With the alignments between items and KG entities, KG is able to profile items and offer complementary information to the interaction data.

\noindent\textbf{Problem Statement}. 
Given the user-item interaction matrix $\mathbf{Y}$ and the knowledge graph $\mathcal{G}$, knowledge-aware recommendation task aims to learn a function that can predict how likely a user would adopt an item.

\section{Methodology}
\begin{figure*}[th]
  \centering
  \includegraphics[width=\textwidth]{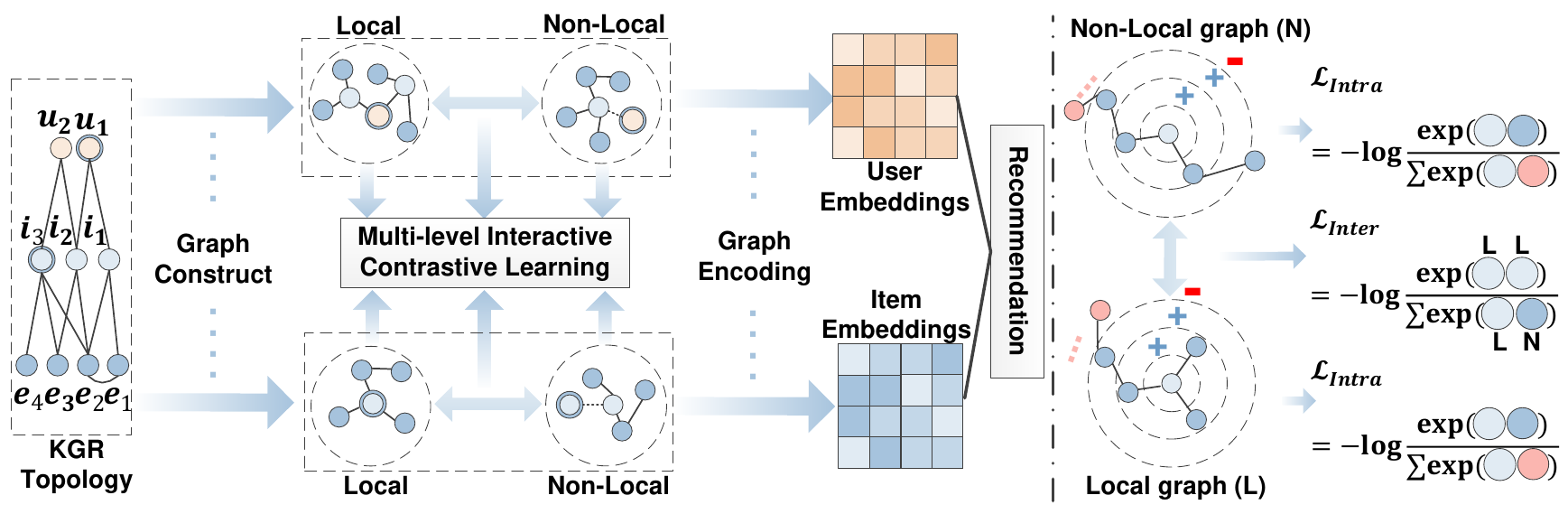}
  \caption{Illustration of the proposed KGIC model. The left subfigure shows model framework of KGIC; and the right subfigure presents the details of Multi-level Interactive Contrastive Learning. Best viewed in color.}
  \label{fig:model}
\end{figure*}

We now present the proposed \underline{K}nowled\underline{g}e-aware Recommender System with Multi-level \underline{I}nteractive \underline{C}ontrastive Learning (KGIC). KGIC aims at unifying the CF and KG parts with an interactive contrastive mechanism for coherent information utilization and hence improving the user/item representation learning.
Figure \ref{fig:model} displays the working flow of KGIC, which mainly consists of three key components: 
1) Graph Constructing and Encoding. It first constructs local and non-local graphs for user/item by combining different CF signals with KG facts, then encodes each layer with a simple GNN encoder in all the graphs. 
2) Intra-graph Interactive Contrastive Learning. It performs interactive contrastive Learning within local and non-local graphs, for unifying the CF and KG information, and further having a consistent and sufficient representation learning.
3) Inter-graph Interactive Contrastive Learning. It performs interactive contrastive Learning between local and non-local graphs, aiming to extract more informative KG signals.
We next present the three components in detail.

\subsection{Graph Constructing and Encoding}
Different from previous KG-aware recommendation methods that only integrate KG entities of local neighboring areas into user/item representation learning, we propose to incorporate the non-local KG information which could be acquired from neighboring KG entities of similar items (\ie co-occurrence items in CF), aiming to fully explore the external facts in KG. 
We first construct local and non-local graphs (\ie the local and non-local sub-KGs) for user/item by combining corresponding CF signals with KG. Then an attentive embedding mechanism is adopted for layer encoding in each graph.

\subsubsection{Local Graph Construction}
Local graph consists of first-order CF (\ie the user's interacted items or the item itself) and related KG facts for user/item. 
Firstly, first-order CF signals for user/item are extracted from the user-item interactions $Y$. Then through the item-entity alignments $\mathcal{A} =\{(v, e)|v \in \mathcal{V}, e \in \mathcal{E}\}$, the items of first-order CF signals are aligned with KG and the initial entities in KG are acquired as follows:
\begin{equation}
\label{initial}
	\begin{array}{ll}
	    \mathcal{E}_{u, L}^{0}=\left\{e \mid(v, e) \in \mathcal{A}, \text { and } v \in\left\{v \mid y_{uv}=1\right\}\right\},	\\
		\mathcal{E}_{v, L}^{0}=\{e \mid(v, e) \in \mathcal{A}\},
	\end{array}
\end{equation}
where $\mathcal{E}_{u, L}^{0}$ and $\mathcal{E}_{v, L}^{0}$ represents the initial entity sets of KG in local graphs, for the user and the item respectively.
After that, more layers' related KG facts are acquired through a natural propagation in KG (\ie propagating along links in KG). And by doing so, the local graphs for user/item are constructed. The triples in local graphs are obtained as follows:
\begin{equation}
	\begin{array}{ll}
		\mathcal{S}_{o, L}^{l}=\left\{(h, r, t) \mid(h, r, t) \in \mathcal{G} \text { and } h \in \mathcal{E}_{o, L}^{l-1}\right\}, l=1, \ldots, L, 
	\end{array}
\end{equation}
where the symbol $o$ is a uniform placeholder which means $u$ or $v$, $\mathcal{S}_{u, L}^{l}$ and $\mathcal{S}_{v, L}^{l}$ represents the triple set in the local graph's $l-$th layer for user and item, which is composed of $(l-1)-$th layer head entity, relation and $l-$th layer tail entity.  
In this way, we obtain a $L-$layer local graph for user/item, which contains user-item-entity and item-entity heterogeneous structure for user and item respectively.

\subsubsection{Non-Local Graph Construction}
Non-local graph contains high-order CF signals (\ie item-user-item co-occurrence items in CF) and more external KG facts for user/item.
Similarly, the non-local graph is constructed by aligning high-order items of CF with KG and KG propagating. Firstly, the high-order items for user/item are acquired through propagating in user-item interactions, as follows:
\begin{equation}
	\begin{array}{lll}
		\mathcal{V}_{p}=\left\{v_{p} \mid u \in \mathcal{U}_{\text {sim }}, \text { and } y_{u v_{p}}=1\right\},	\\
		\mathcal{U}_{\text {sim }}=\left\{u_{\text {sim }} \mid v \in\left\{v \mid y_{u v}=1\right\} \text { and } y_{u_{\text {sim }} v}=1\right\},	\\
		\mathcal{V}_{u}=\left\{v_{u} \mid u \in\left\{u \mid y_{u v}=1\right\} \text { and } y_{u v_{u}}=1\right\},
	\end{array}
\end{equation}
where $\mathcal{V}_{p}$ and $\mathcal{U}_{\text {sim }}$ represent high-order items and similar users of the user, and $\mathcal{V}_{u}$ is the high-order items of the item.
Further the aligned initial entities $\mathcal{E}_{u, N}^{0}$ and $\mathcal{E}_{v, N}^{0}$ in KG are acquired as follows:
\begin{equation}
	\begin{array}{lll}
	    \mathcal{E}_{u, N}^{0}=\left\{e \mid\left(v_{p}, e\right) \in \mathcal{A}, \text { and } v_{p} \in \mathcal{V}_{p}\right\},	\\
    	\mathcal{E}_{v, N}^{0}=\left\{e \mid\left(v_{u}, e\right) \in \mathcal{A}, \text { and } v_{u} \in \mathcal{V}_{u}\right\}.
	\end{array}
\end{equation}

Then by propagating these initial entities in KG, the non-local graph for user/item is constructed, whose triples are formed as follows:
\begin{equation}
	\begin{array}{ll}
	   \mathcal{S}_{o, N}^{l}=\left\{(h, r, t) \mid(h, r, t) \in \mathcal{G} \text {and} h \in \mathcal{E}_{o, N}^{l-1}\right\}, l=1, \ldots, L.
	\end{array}
\end{equation}

\subsubsection{Graph Encoding}
Obtaining local and non-local graphs of user/item, it's indispensable to encode each layer information into an embedding for learning a comprehensive representation. Inspired by previous KG-aware recommendation works, an attentive embedding mechanism is adopted here, which reveals the different meanings of the tail entities with different contexts and generates different attentive weights.
Considering $\left(h, r, t\right)$ the $i$-th triple of the $l$-th layer triple set, we can get the representation of the $l$-th layer as follows:

\begin{align}
\Vec{E}_{o, D}^{l}=\sum_{i=1}^{m} \pi\left(e_{i}^{h}, r_{i}\right) e_{i}^{t},
\end{align}
where the symbol $D$ is a uniform placeholder which means $L$ or $N$, $\Vec{E}_{o, D}^{l}$ represents $l$-th layer’s embedding of $u$ or $v$ in local/non-local graph. And $m$ is the number of triples in the $l$-th layer, $e_{i}^{h}$, $r_{i}$ and $e_{i}^{t}$ are the embeddings of head entity, relation and the tail entity of the $i$-th triple. The weights $\pi(e_{i}^{h}, r_{i})$ are acquired by the attentive mechanism which can be described as follows:
\begin{equation}
    \label{attentive_mechanism}
	\begin{split}
	\pi\left(e_{i}^{h}, r_{i}\right)&=\sigma\left(W_{1}\left[\sigma\left(W_{0}\left(e_{i}^{h}|| r_{i}\right)+b_{0}\right)\right]+b_{1}\right),	\\
	\pi\left(\mathrm{e}_{i}^{h}, \mathrm{r}_{i}\right)&=\frac{\exp \left(\pi\left(\mathrm{e}_{i}^{h}, \mathrm{r}_{i}\right)\right)}{\sum_{\left(h^{\prime}, r^{\prime}, t^{\prime}\right) \in \mathcal{S}_{o, D}^{l}} \exp \left(\pi\left(\mathrm{e}_{i}^{h^{\prime}}, \mathrm{r}_{i}^{\prime}\right)\right)},	
	\end{split}
\end{equation}
where $||$ is the concatenation operation, $W_{*} \in \mathbf{R}^{2d \times d}$ and $b_{*} \in \mathbf{R}^{d}$ are the trainable weight matrices and biases. Hence we successfully encode each layer of local and non-local graphs into embeddings. 

\subsection{Intra-graph Interactive Contrastive Learning}
Based on the constructed local and non-local graphs of user/item, we move on to encourage a balanced information utilization in these graphs. Since the local and non-local graphs have a relatively unbalanced heterogeneous structure, which is composed of sparse user-item interactions and redundant KG connections, thus the crucial CF signals tend to have less impact on representation learning. The intra-graph interactive contrastive learning is hence proposed to make coherent use of CF and KG, by performing interactions between CF and KG information with contrastive learning.

Specifically, the intra-graph interactive contrastive learning regards the CF part (\ie the initial entities in Equation~\eqref{initial} ) as the anchor, which is located in center layer of the local/non-local graph. The knowledge information involved in aggregation for learning user/item representation forms the positive pairs, and the other KG entities (\ie the unemployed higher layers in graphs) form the negative pairs.
With the defined positive and negative pairs, we have the following contrastive loss for the user:

\begin{equation}\label{intra_loss}
\begin{array}{ll}
\mathcal{L}_{Intra}^{U}&=\sum\limits_{u \in \mathcal{U}} -\log \frac{\sum\limits_{k \in L}e^{\left(\left(\Vec{E}_{u, L}^{(0)} \cdot \Vec{E}_{u, L}^{(k)} / \tau\right)\right)}}{\underbrace{\sum\limits_{k \in L}e^{\left(\left(\Vec{E}_{u, L}^{(0)} \cdot \Vec{E}_{u, L}^{(k)} / \tau\right)\right)}}_{\text {positive pair}} + \underbrace{\sum\limits_{k' > L} e^{\left(\left(\Vec{E}_{u, L}^{(0)} \cdot \Vec{E}_{u, L}^{(k')} / \tau\right)\right)}}_{\text {intra-graph negative pair}}}   \\

&+ \sum\limits_{u \in \mathcal{U}} -\log \frac{\sum\limits_{k \in L}e^{\left(\left(\Vec{E}_{u, N}^{(0)} \cdot \Vec{E}_{u, N}^{(k)} / \tau\right)\right)}}{\underbrace{\sum\limits_{k \in L}e^{\left(\left(\Vec{E}_{u, N}^{(0)} \cdot \Vec{E}_{u, N}^{(k)} / \tau\right)\right)}}_{\text {positive pair}} + \underbrace{\sum\limits_{k' > L} e^{\left(\left(\Vec{E}_{u, N}^{(0)} \cdot \Vec{E}_{u, N}^{(k')} / \tau\right)\right)}}_{\text {intra-graph negative pair}}},
\end{array}
\end{equation}
where $\tau$ denotes a temperature parameter in softmax. In a similar way, the intra-graph contrastive loss of the item can be obtained as $\mathcal{L}_{Intra}^{I}$.
And the complete Intra-graph interactive contrastive Loss is the sum of the above two losses:
\begin{equation}
    \mathcal{L}_{Intra}= \mathcal{L}_{Intra}^{U}+ \mathcal{L}_{Intra}^{I}.
\end{equation}

In this way, CF and KG signals successfully supervise each other to have a more coherent and sufficient representation learning.

\subsection{Inter-graph Interactive Contrastive Learning}
Although intra-graph interactive contrastive learning has achieved coherent information utilization in each single graph, it's still a challenge to integrate the local and non-local information together for that the non-local one is noisier.
Since the non-local graph is composed of high-order CF signals and its corresponding KG facts, more external useful external facts along with noisy information are included. Hence the inter-graph interactive contrastive learning is proposed, to extract informative non-local information by contrasting the non-local graph with the cleaner local graph and collaboratively supervising each other.

To be more specific, the inter-graph interactive contrastive learning treats any layers of the local graph as the anchor, and the same layer in the non-local graph forms the positive pair, and the other layers in the non-local graph are naturally regarded as the negative pairs. Similar to intra-graph interactive contrastive mechanism, we have the following contrastive loss for the user:
\begin{equation}\label{inter_loss}
\begin{array}{ll}
\mathcal{L}_{Inter}^{U}=\sum\limits_{u \in \mathcal{U}} \sum\limits_{k \in L}-\log \frac{e^{\left(\left(\Vec{E}_{u, L}^{(k)} \cdot \Vec{E}_{u, N}^{(k)} / \tau\right)\right)}}{\underbrace{e^{\left(\left(\Vec{E}_{u, L}^{(k)} \cdot \Vec{E}_{u, N}^{(k)} / \tau\right)\right)}}_{\text {positive pair}} + \underbrace{\sum\limits_{k' \neq k} e^{\left(\left(\Vec{E}_{u, L}^{(k)} \cdot \Vec{E}_{u, N}^{(k')} / \tau\right)\right)}}_{\text {inter-graph negative pair}}}.
\end{array}
\end{equation}

Similarly, we could obtain the inter-graph contrastive loss of the item $\mathcal{L}_{Inter}^{I}$. Then the complete inter-graph contrastive loss is the sum of the user and item contrastive loss:
\begin{equation}
    \mathcal{L}_{Inter}= \mathcal{L}_{Inter}^{U} + \mathcal{L}_{Inter}^{I}.
\end{equation}

\subsection{Model Prediction}
Having updated all the layers' embeddings, we obtain every layer's embedding for user's local and non-local graph, item’s local and non-local graph. Every layer’s embedding expresses a part of user/item representation and stresses the influence of different layers and components. By concatenating these representation vectors of each layer,
final user/item embedding is concluded for predicting their matching score through inner product, as follows:
\begin{equation}
\begin{split} 
\Vec{e}_{u} &=\Vec{E}_{u, L}^{0} \|\ldots\| \Vec{E}_{u, L}^{L} \| \Vec{E}_{u, N}^{0} \|\ldots\| \Vec{E}_{u, N}^{L}, \\
\Vec{e}_{i} &=\Vec{E}_{i, L}^{0} \|\ldots\| \Vec{E}_{i, L}^{L} \| \Vec{E}_{i, N}^{0} \|\ldots\| \Vec{E}_{i, N}^{L}, \\
\hat{y}(u, i) &=\Vec{e}_{u}^{\top} \Vec{e}_{i}.
\end{split}
\end{equation}

\subsection{Multi-task Training}
To combine the recommendation task with the self-supervised task, we
treat the proposed two contrastive learning losses as supplementary. Hence a multi-task learning strategy is leveraged to jointly train the KG-aware recommendation loss and the proposed contrastive loss.
For the KG-aware recommendation task, a pairwise BPR loss \cite{rendle2012bpr} is adopted to reconstruct the historical data, which encourages the prediction scores of a user’s historical items to be higher than the unobserved items.
\begin{equation}
    \mathcal{L}_{\mathrm{BPR}}=\sum_{(u, i, j) \in O}-\ln \sigma\left(\hat{y}_{u i}-\hat{y}_{u j}\right),
\end{equation}
where $\boldsymbol{O}=\left\{(u, i, j) \mid(u, i) \in \boldsymbol{O}^{+},(u, j) \in \boldsymbol{O}^{-}\right\}$ is the training dataset consisting of the observed interactions $\boldsymbol{O}^{+}$ and unobserved counterparts $\boldsymbol{O}^{-}$; $\sigma$ is the sigmoid function. 
By combining the intra- and inter-graph interactive contrastive loss with BPR loss, we minimize the following objective function to learn the model parameter:
\begin{equation}\label{overall_loss}
     \mathcal{L}_{KGIC} = \mathcal{L}_{\mathrm{BPR}} + \lambda1(\alpha\mathcal{L}_{Intra} + \mathcal{L}_{Inter}) + \lambda2\|\Theta\|_{2}^{2},
\end{equation}
where $\Theta$ is the model parameter set, $\alpha$ is the hyperparameter to balance the weight of the intra- and inter-graph contrastive losses, $\lambda1$ and $\lambda2$ are two hyperparameters to control the contrastive loss and $L_2$ regularization term, respectively.
\section{Experiment}

\begin{table}[tb]
\centering
\setlength{\tabcolsep}{0.4mm}{
\begin{tabular}{cl|ccc}
\hline
\multicolumn{1}{l}{}                                                                &                 & \multicolumn{1}{l}{Book-Crossing} & \multicolumn{1}{l}{MovieLens-1M} & \multicolumn{1}{l}{Last.FM} \\ \hline \hline
\multirow{3}{*}{\begin{tabular}[c]{@{}c@{}}User-item   \\ Interaction\end{tabular}} & \# users  & 17,860 & 6,036 & 1,872                       \\
& \# items & 14,967 & 2,445 & 3,846                       \\
& \# interactions & 139,746 & 753,772 & 42,346                      \\ \hline
\multirow{3}{*}{\begin{tabular}[c]{@{}c@{}}Knowledge\\ Graph\end{tabular}} & \# entities & 77,903 & 182,011 & 9,366                       \\
& \# relations & 25 & 12 & 60                          \\
& \# triplets & 151,500 & 1,241,996 & 15,518                      \\ \hline
\multirow{3}{*}{\begin{tabular}[c]{@{}c@{}}Hyper-\\parameter\\ Settings\end{tabular}} 

& \# $\eta$  & $4 \times 10^{-3}$ & $4 \times 10^{-3}$ & $4 \times 10^{-3}$                          \\
& \# $\lambda1$  & $1 \times 10^{-6}$ & $1 \times 10^{-7}$ & $1 \times 10^{-6}$                      \\
& \# $\lambda2$  & $1 \times 10^{-4}$ & $1 \times 10^{-5}$ & $1 \times 10^{-4}$                      \\
\hline
\end{tabular}}
\caption{Statistics and hyper-parameter settings for the three datasets. ($\eta$: learning rate, $\lambda1$: constrastive loss weight, $\lambda2$: L2 regularizer weight.)}
\label{tab:datasets}
\end{table}

%



Extensive experiments have been done on three public datasets, for answering the following research questions:

\begin{itemize}[leftmargin=*, nosep]
    \item \textbf{RQ1}: How does KGIC perform, compared with the state-of-the-art models?
    \item \textbf{RQ2}: How do the main components (\eg intra- and inter-graph interactive contrastive learning) affect KGIC performance?
    \item \textbf{RQ3}: How do different hyper-parameter settings (\eg model depth, coefficient $\alpha$, temperature $\tau$) affect KGIC?
    \item \textbf{RQ4}: Is the self-supervised task really improving the representation learning quality?
\end{itemize}

\subsection{Experiment Settings}

\subsubsection{Dataset Description.}
Three publicly available datasets are used to evaluate the effectiveness of KGIC: Book-Crossing, MovieLens-1M, and Last.FM, which vary in size and sparsity, making our experiments more convincing. The basic statistics of the three datasets are presented in Table~\ref{tab:datasets}.
\begin{itemize}[leftmargin=*, nosep]
\item{\verb|Book-Crossing|\footnote{\url{http://www2.informatik.uni-freiburg.de/~cziegler/BX/}}}: It consists of trenchant ratings (ranging from 0 to 10) about various books from the book-crossing community.
\item {\verb|MovieLens-1M|\footnote{\url{https://grouplens.org/datasets/movielens/1m/}}}: It contains approximately 1 million explicit ratings (ranging from 1 to 5) for movie recommendations.
\item{\verb|Last.FM|\footnote{\url{https://grouplens.org/datasets/hetrec-2011/}}}: It is collected from Last.FM online music systems, containing listening history with around 2 thousand users.
\end{itemize}

Note that we follow RippleNet \cite{wang2018ripplenet} to transform the explicit feedback in three datasets into the implicit one where 1 indicates the positive samples. For negative samples, we randomly sample unobserved items with the same size as positive ones for each user.
As for the sub-KG construction, we use Microsoft Satori\footnote{\url{https://searchengineland.com/library/bing/bing-satori}}, closely following RippleNet \cite{wang2018ripplenet} and KGCN\cite{wang2019knowledge}. Each sub-KG follows the triple format and is a subset of the whole KG with a confidence level over 0.9. Given the sub-KG, we gather Satori IDs of all valid movies/books/musicians through matching their names with the tail of triples. Then we match the item IDs with the head of all triples and select all well-matched triples from the sub-KG. 

\subsubsection{Evaluation Metrics.}
To comprehensively evaluate our method, we conduct the evaluation in two experimental scenarios: (1) In click-through rate (CTR) prediction, two widely used metrics \cite{wang2018ripplenet, wang2019knowledge} $AUC$ and $F1$ are adopted here. (2) In top-$K$ recommendation, we choose Recall@$K$ to evaluate the recommended sets, where $K$ is set to 5, 10, 20, 50, and 100 for consistency.

\subsubsection{Baselines.}
To demonstrate the effectiveness of our proposed KGIC, we compare KGIC with 
the state-of-the-art methods, covering: CF-based methods (BPRMF), embedding-based methoda (CKE, RippleNet), path-based methoda (PER), and GNN-based methods (KGCN, KGNN-LS, KGAT, CKAN, KGIN, CG-KGR) as follows:
\begin{itemize}[leftmargin=*, nosep]
\item {\verb|BPRMF| \cite{rendle2012bpr}}: It’s a typical CF-based method that uses pairwise matrix factorization for implicit feedback optimized by BPR loss.
\item{\verb|CKE| \cite{zhang2016collaborative}}: It’s an embedding-based method that combines structural, textual, and visual knowledge in one framework.
\item {\verb|RippleNet| \cite{wang2018ripplenet}}: It’s a classical embedding-based method which propagates users’ preferences on the KG.
\item{\verb|PER| \cite{yu2014personalized}}: It’s a path-based method which extracts meta-path features to represent the connectivity between users and items.
\item {\verb|KGCN| \cite{wang2019knowledge}}: It’s a GNN-based method which iteratively integrates neighboring information to enrich item embeddings.
\item {\verb|KGNN-LS| \cite{wang2019knowledge-aware}}: It is a GNN-based model which enriches item embeddings with GNN and label smoothness regularization.
\item {\verb|KGAT| \cite{wang2019kgat}}: It’s a GNN-based method which iteratively integrates neighbors on user-item-entity graph with an attention mechanism to get user/item representations.
\item {\verb|CKAN| \cite{wang2020ckan}}: It’s a GNN-based method which independently propagates collaborative and knowledge signals on CF and KG parts.
\item {\verb|KGIN| \cite{wang2021learning}}: It's a state-of-the-art GNN-based method, which disentangles user-item interactions at the granularity of user intents, and performs GNN on the user-intent-item-entity graph.
\item {\verb|CG-KGR| \cite{chen2022attentive}}: It’s the latest GNN-based method which fuses the collaborative signals into knowledge aggregation with GNN. 
\end{itemize}

\subsubsection{Parameter Settings.}
We implement our KGIC and all baselines in Pytorch and carefully tune the key parameters. We fix the embedding size to 64 in all models for a fair comparison. The default Xavier method \cite{glorot2010understanding} is adopted to initialize the model parameters. Besides, we utilize Adam \cite{kingma2014adam} optimizer and set the batch size to 2048. The local and non-local triple set size are limited to 40 and 128 respectively.
Other hyper-parameter settings are provided in Table~\ref{tab:datasets}. The best settings for hyperparameters in all methods are researched by either empirical study or the original papers. 

\begin{table*}[tb]
    \centering
    \setlength{\tabcolsep}{8pt}
    \begin{tabular}{c|ll|ll|ll}
    	\hline
    	\multirow{2}{*}{Model} & \multicolumn{2}{c}{Book-Crossing} & \multicolumn{2}{c}{MovieLens-1M} & \multicolumn{2}{c}{Last.FM} \\
        & \multicolumn{1}{c}{\textit{AUC}} & \multicolumn{1}{c}{\textit{F1}} & \multicolumn{1}{c}{\textit{AUC}} & \multicolumn{1}{c}{\textit{F1}} & \multicolumn{1}{c}{\textit{AUC}} & \multicolumn{1}{c}{\textit{F1}} \\
        \hline\hline
BPRMF & 0.6583$(-11.66\%)$ & 0.6117$(-6.95\%)$ & 0.8920$(-3.32\%)$ & 0.7921$(-6.38\%)$ & 0.7563$(-10.29\%)$ & 0.7010$(-7.43\%)$    \\
\hline
CKE   & 0.6759$(-9.90\%)$ & 0.6235$(-5.77\%)$ & 0.9065$(-1.87\%)$ & 0.8024$(-5.35\%)$ & 0.7471$(-11.21\%)$ & 0.6740$(-10.13\%)$  \\
RippleNet   & 0.7211$( -5.38\% )$ & 0.6472$( -3.40\% )$ & 0.9190$(-0.62\% )$ & 0.8422$( -1.37\% )$ & 0.7762$( -8.30\% )$ & 0.7025$( -7.28\% )$ \\
\hline
PER   & 0.6048$(-17.01\%)$ & 0.5726$(-10.86\%)$ & 0.7124$(-21.28\%)$ & 0.6670$(-18.89\%)$ & 0.6414$(-21.78\%)$ & 0.6033$(-17.20\%)$          \\
\hline
KGCN     & 0.6841$(-9.08\%)$ & 0.6313$(-4.99\%)$  & 0.9090$(-1.62\%)$ & 0.8366$(-1.93\%)$ & 0.8027$(-5.65\%)$ & 0.7086$(-6.67\%)$          \\
KGNN-LS    & 0.6762$(-9.87\%)$ & 0.6314$(-4.98\%)$ & 0.9140$(-1.12\%)$ & 0.8410$(-1.49\%)$ & 0.8052$(-5.40\%)$ & 0.7224$(-5.29\%)$          \\
KGAT    & 0.7314$(-4.35\%)$ & 0.6544$(-2.68\%)$ & 0.9140$(-1.12\%)$ & 0.8440$(-1.19\%)$ & 0.8293$(-2.99\%)$ & 0.7424$(-3.29\%)$          \\
CKAN    & 0.7420$(-3.29\%)$ & 0.6671$(-1.41\%)$ & 0.9082$(-1.70\%)$ & 0.8410$(-1.49\%)$ & 0.8418$(-1.74\%)$ & 0.7592$(-1.61\%)$      \\
KGIN    & 0.7273$(-4.76\%)$   & 0.6614$(-1.98\%)$ & \underline{0.9190}$(-0.62\%)$ & \underline{0.8441}$(-1.18\%)$ & \underline{0.8486}$(-1.06\%)$ & \underline{0.7602}$(-1.51\%)$ \\
CG-KGR    & \underline{0.7498}$(-2.51\%)$ & \underline{0.6689}$(-1.23\%)$ & 0.9110$(-1.42\%)$ & 0.8359$(-2.00\%)$ & 0.8336$(-2.56\%)$ & 0.7433$(-3.20\%)$      \\
 \hline
\textbf{KGIC}    & \textbf{0.7749}* & \textbf{0.6812}* & \textbf{0.9252}* & \textbf{0.8559}* & \textbf{0.8592}* & \textbf{0.7753}*           \\ \hline
\end{tabular}
\caption{The result of $AUC$ and $F1$ in CTR prediction. The best results are in boldface and the second best results are underlined. * denotes statistically significant improvement by unpaired two-sample $t$-test with $p < 0.001$.}
\label{tab:compare}
\vspace{-0.6cm}
\end{table*}

\begin{figure*}[htb]  
    \centering  
    \subfloat[Book-Crossing] 
    {
        \begin{minipage}[t]{0.3\textwidth}
            \centering     
            \includegraphics[height=0.8\textwidth]{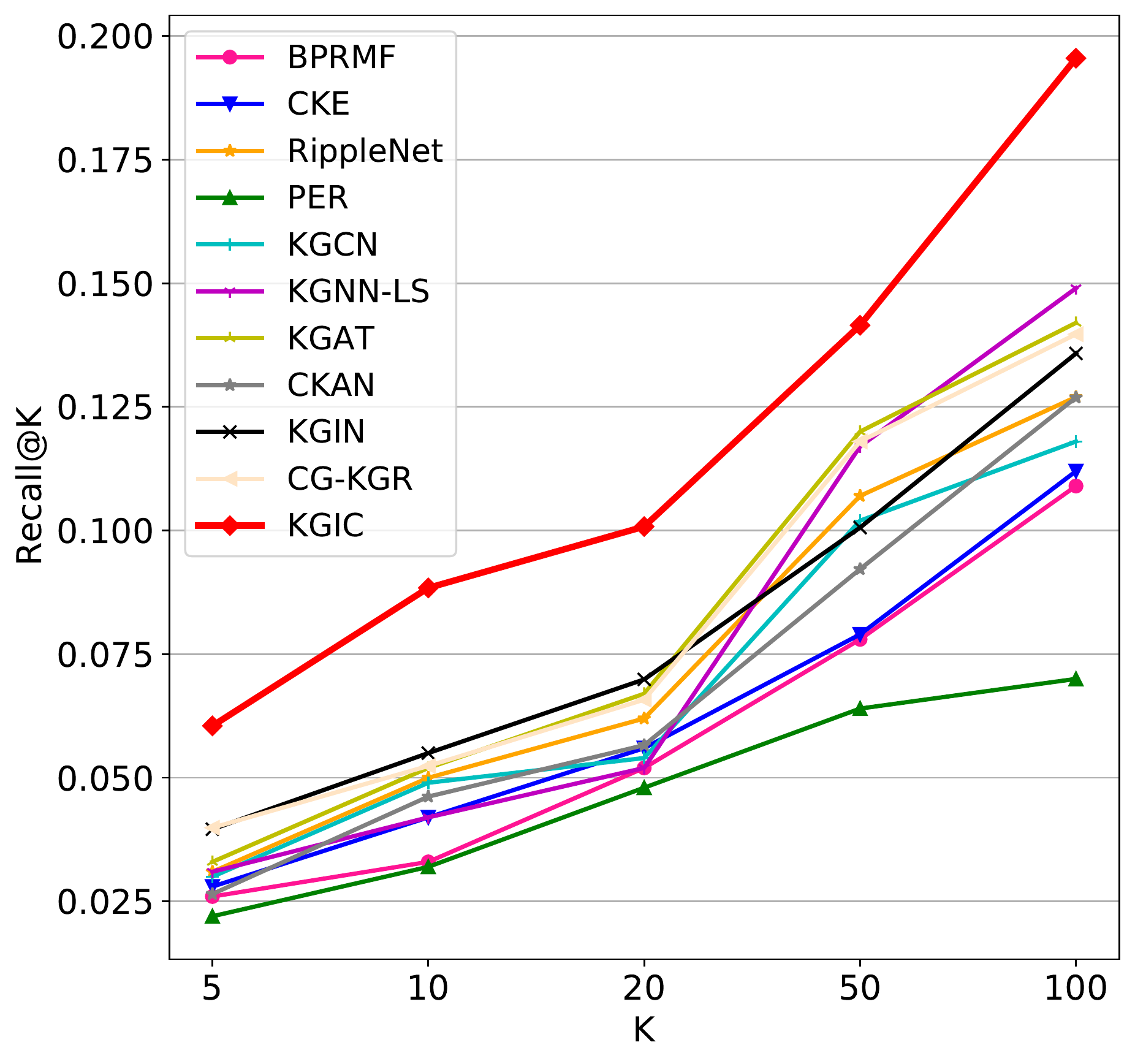}   
        \end{minipage}
    }
    \subfloat[MovieLens-1M] 
    {
        \begin{minipage}[t]{0.3\textwidth}
            \centering     
            \includegraphics[height=0.8\textwidth]{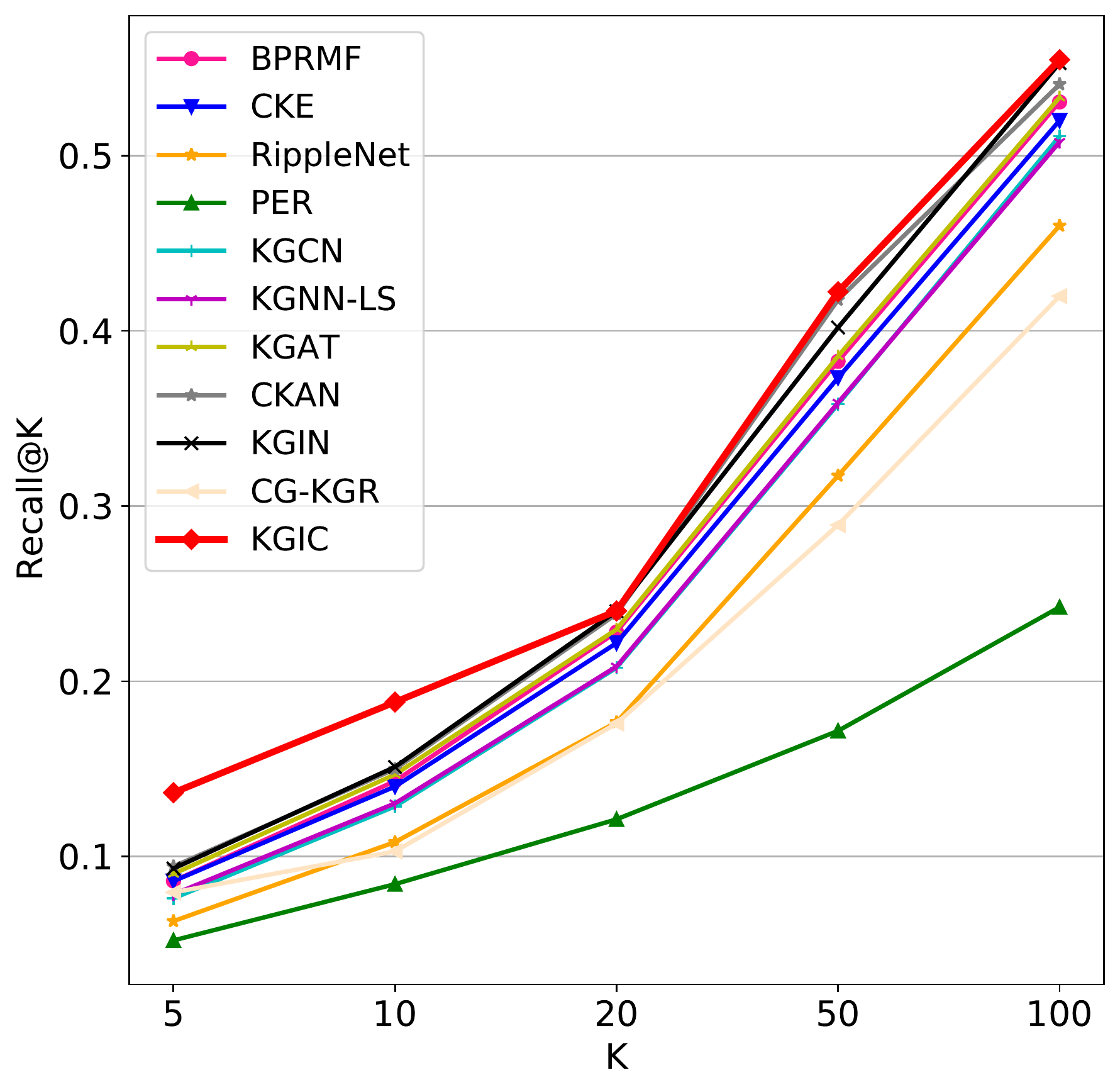}   
        \end{minipage}
    }
    \subfloat[Last.FM] 
    {
        \begin{minipage}[t]{0.3\textwidth}
            \centering      
            \includegraphics[height=0.8\textwidth]{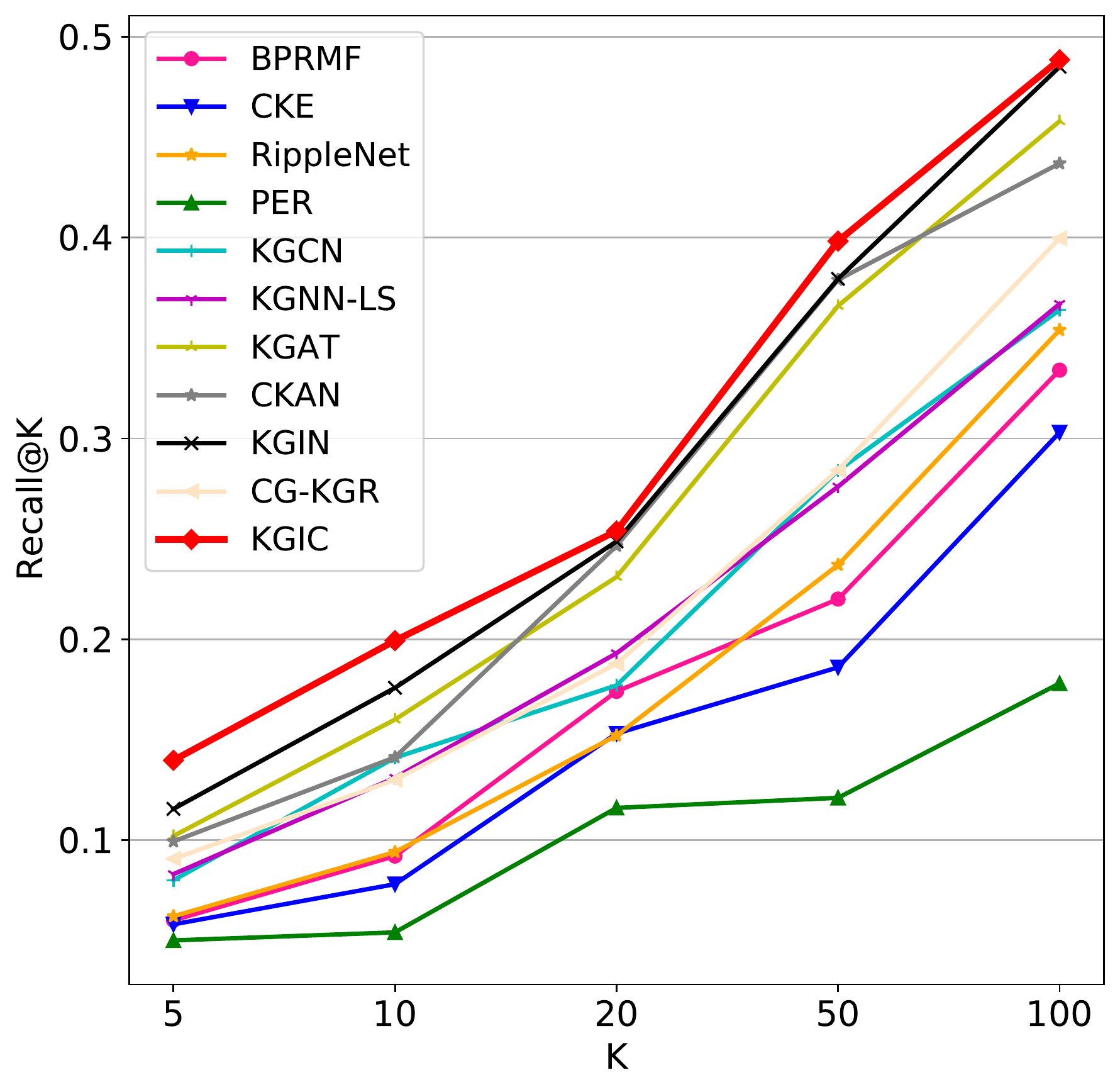}   
        \end{minipage}
    }%
    \caption{The result of Recall@$K$ in top-$K$ recommendation.} 
    \label{fig:topk} 
\end{figure*}

\subsection{Performance Comparison (RQ1)}
The empirical results of all methods in two scenarios are reported in Table~\ref{tab:compare} and Figure~\ref{fig:topk}, respectively. The improvements and statistical significance test are performed between KGIC and the strongest baselines (highlighted with underline). By analyzing the performance comparison, we have the following observations:

\begin{itemize}[leftmargin=*, nosep]
    \item \textbf{Our proposed KGIC achieves the best results.} KGIC yields the best performance across three datasets in terms of all measures. More specifically, KGIC improves \wrt $AUC$ by 2.51\%, 0.62\%, and 1.06\% over the strongest baselines in Book-Crossing, MovieLens-1M, and Last.FM datasets respectively. In top-$K$ recommendation scenario, KGIC also performs best \wrt $Recall@K$ (k = 5, 10, 20, 50, 100) in all the cases. We attribute such improvements to the following aspects:
    (1) By contrasting the CF with KG signals in local/non-local graphs, the intra-graph level interactive contrastive learning performs interactions between two parts and supervises each other to improve the representation learning.
    (2) Through contrasting the local and non-local graphs of user/item, inter-graph level interactive contrastive learning sufficiently incorporates the non-local KG facts and learns discriminative representations from the two kinds of graphs.
    \item \textbf{Most KG-aware models achieve better performance.} We can observe that models with KG mostly perform better than conventional CF methods. Comparing CKE with BPRMF, simply incorporating KG embeddings into MF boosts the model performance. This confirms the importance of bringing in KG.
    \item \textbf{Simply integrating KG is not a guarantee of performance improvement.} Traditional CF-based methods BPRMF works slightly better than embedding-based method CKE and path-based method PER in some scenarios. This phenomenon reveals that the overly unbalanced utilization of KG would unexpectedly degrade the model performance, which stresses the importance of making sufficient and coherent use of KG.
    \item \textbf{Extracting more informative KG facts boosts the model performance} GNN-based methods have a better performance than embedding-based and path-based ones in most cases, which indicates the effectiveness of extracting long-range KG facts. This fact convinces that exploring more informative KG facts highly related to user/item would facilitate the representation, which motivates us to explore more knowledge entities ignoring the limitation of local areas.
\end{itemize}

\subsection{Ablation Studies (RQ2)}

\begin{figure*}[htb] 
    \centering  
    \includegraphics[width=\linewidth]{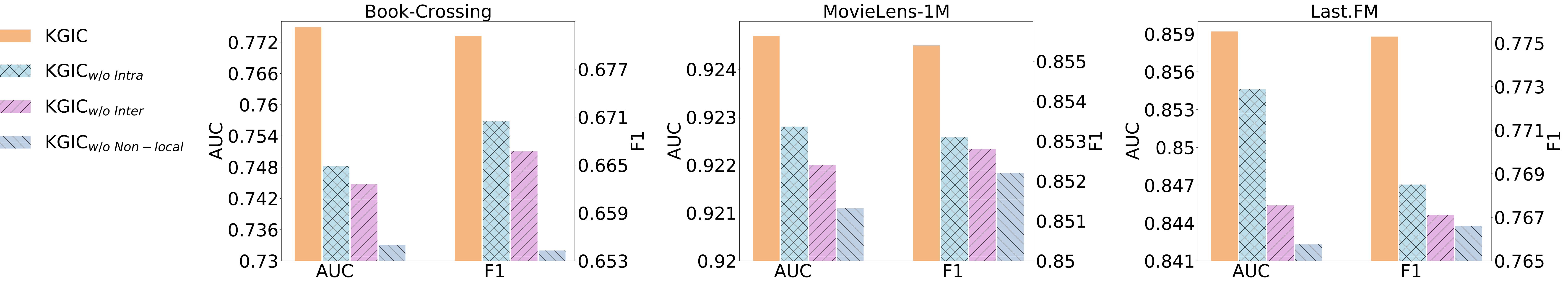}
    \vspace{-0.4cm}
    \caption{Effect of ablation study.} 
    \label{fig:ablation} 
\end{figure*}

Towards examining the contributions of main components in our model, we compare the KGIC with the following variants: (1) $KGIC_{w/o \ intra}$: In this variant, we disable the intra-graph interactive contrastive learning module. (2) $KGIC_{w/o \ inter}$: This variant removes inter-graph interactive contrastive learning module. (3) $KGIC_{w/o \ Non-local}$ This variant removes non-local graphs, hence the intra-graph constrastive loss of non-local graph and inter-graph contrastive loss are both disabled. 
The experimental results are reported in Figure~\ref{fig:ablation}, from which we could summarize the following observations:
\begin{itemize}[leftmargin=*, nosep]
    \item Compared with KGIC, $KGIC_{w/o \ intra}$ and $KGIC_{w/o \ inter}$ consistently get a worse performance, which means removing each of the contrastive loss leads to a performance decrease. This fact demonstrates the importance of exploiting interactive contrastive learning to improve user/item representation learning. Besides, it reveals that these two contrastive losses complement each other and improve the performance in different aspects.
    \item Removing Non-local graph significantly degrades the model performance, and $KGIC_{w/o \ Non-local}$ is the least competitive model. It makes sense since $KGIC_{w/o \ Non-local}$ only considers the intra-graph contrastive loss in local graphs, which also verifies the importance of both intra- and inter-graph contrastive learning.
    \item In most cases, all of the three variants surpass the baseline models across the three datasets, which further demonstrates the effectiveness of intra- and inter-graph contrastive learning.
\end{itemize}

\subsection{Sensitivity Analysis (RQ3)}
\subsubsection{Impact of model depth.}

The model depth $L$ represents the aggregation layer in the local/non-local graph, and also represents the layers of positive pairs in the interactive contrastive mechanism.
To study the influence of model depth, we vary $L$ in range of \{1, 2, 3\} and demonstrate the performance comparison on book, movie, and music datasets in Table~\ref{tab::pos_layer}. KGIC performs best when $L=1, 2, 2$, on Book, Movie, and Music respectively. We could observe that: (1) One or two layers are the proper distance for aggregating neighboring information in the local/non-local graph, further stacking more layers only introduces more noise. (2) One or two layers' positive pairs are enough for learning discriminative embeddings while performing interactions between layers of CF and KG.
\subsubsection{Impact of intra-graph contrastive loss weight $\alpha$.}

\begin{table}[ht]
\centering
\setlength{\tabcolsep}{3pt}{
\begin{tabular}{l|c c|c c|c c}
\hline
 & \multicolumn{2}{c|}{Book} & \multicolumn{2}{c|}{Movie} & \multicolumn{2}{c}{Music} \\ 
 & Auc & F1 & Auc & F1 & Auc & F1 \\ \hline\hline
$L$=1 & \textbf{0.7749}	&\textbf{0.6812} & 0.9241  &0.8551 & 0.8482	&0.7692 \\ 
$L$=2 & 0.7689	&0.6705 & \textbf{0.9252}	&\textbf{0.8559} & \textbf{0.8592}	&\textbf{0.7753} \\ 
$L$=3 & 0.7513	&0.6718  & 0.9203 &0.8521  & 0.8511 &0.7694 \\ 
\hline
\end{tabular}}
\caption{Impact of model depth.}
\label{tab::pos_layer}
\end{table}

\begin{figure}[tb] 
    \centering  
    \subfloat[Book]
    {   \begin{minipage}[t]{0.35\linewidth}
            \centering          
            \includegraphics[width=\textwidth]{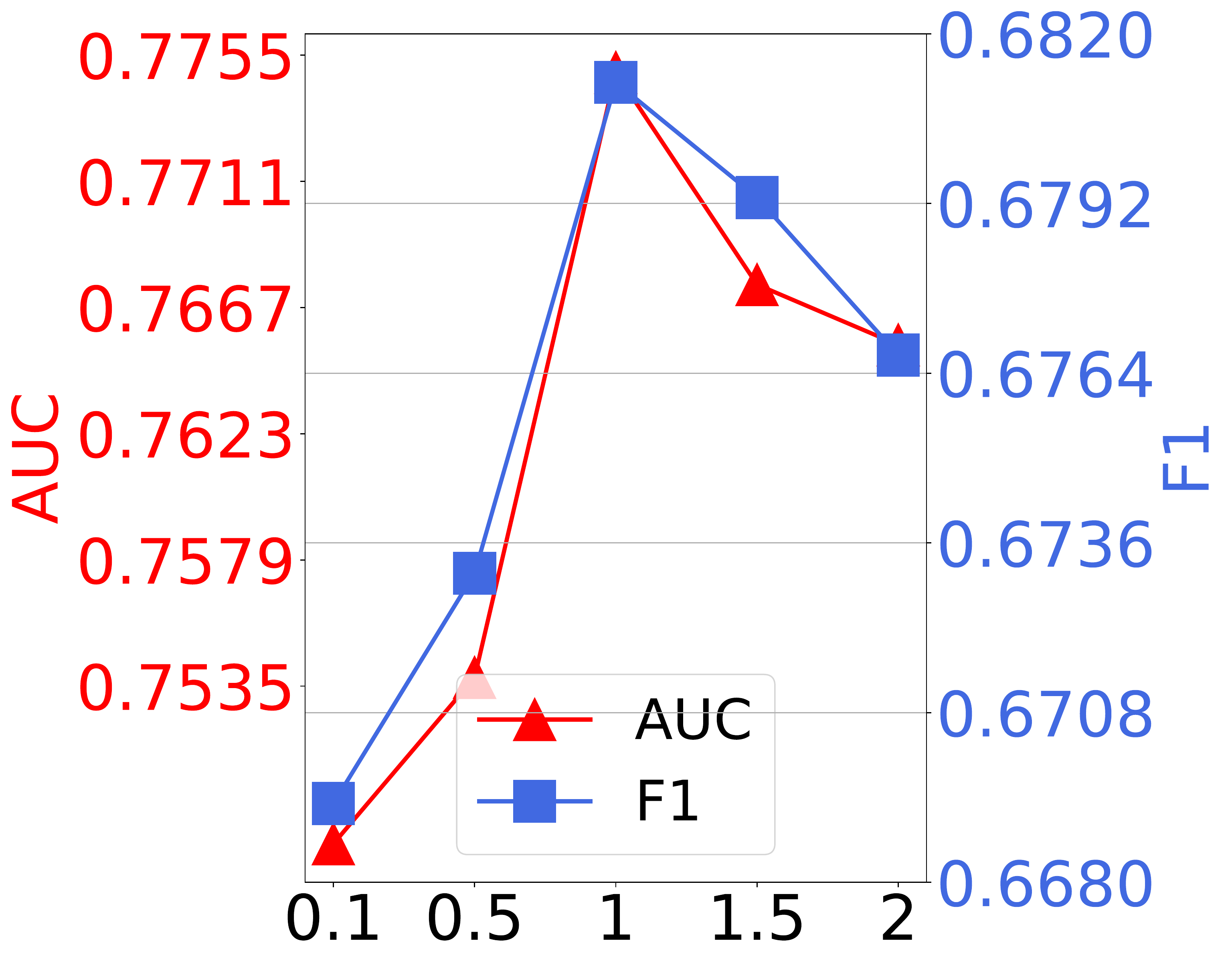} 
        \end{minipage}}
    \subfloat[Movie]
    {   \begin{minipage}[t]{0.35\linewidth}
            \centering     
            \includegraphics[width=\textwidth]{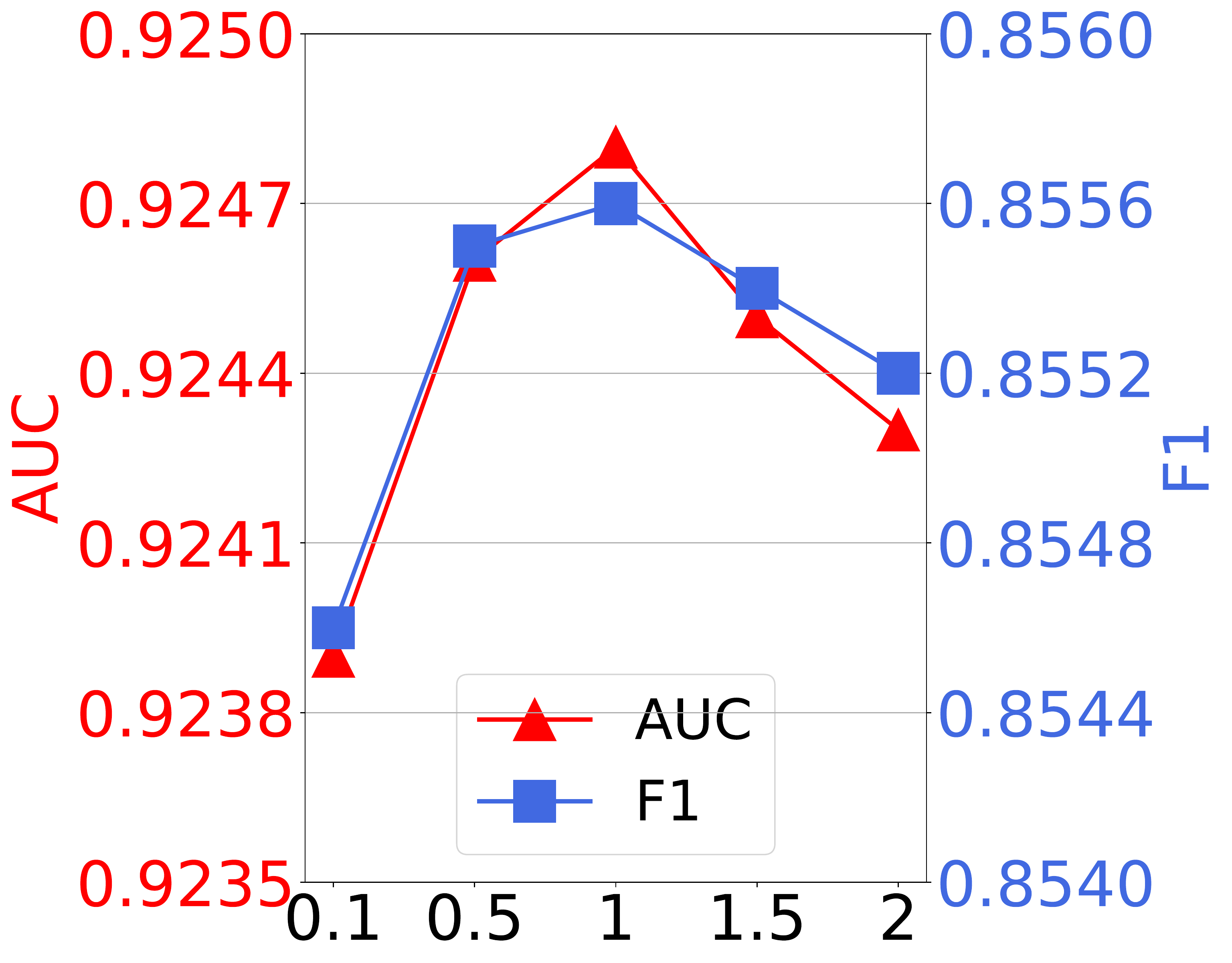} 
        \end{minipage}}%
    \subfloat[Music]
    {   \begin{minipage}[t]{0.35\linewidth}
            \centering      
            \includegraphics[width=\textwidth]{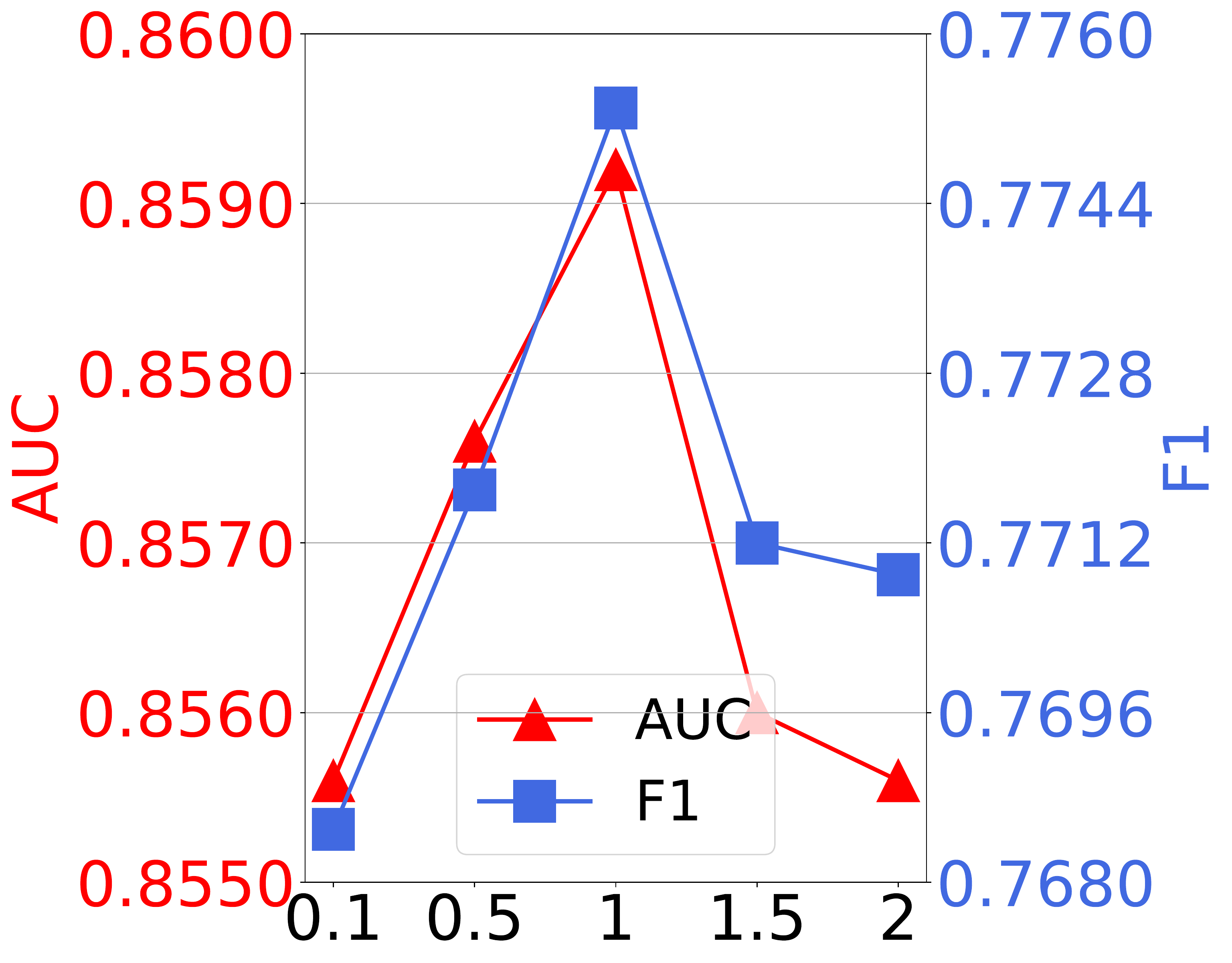}
        \end{minipage}}%
    \vspace{-0.2cm}
    \caption{Impact of coefficient $\alpha$.} 
    \label{fig:alpha} 
\end{figure}

\begin{figure}[tb] 
    \centering  
    \subfloat[Book]
    {   \begin{minipage}[t]{0.5\linewidth}
            \centering          
            \includegraphics[width=0.8\textwidth]{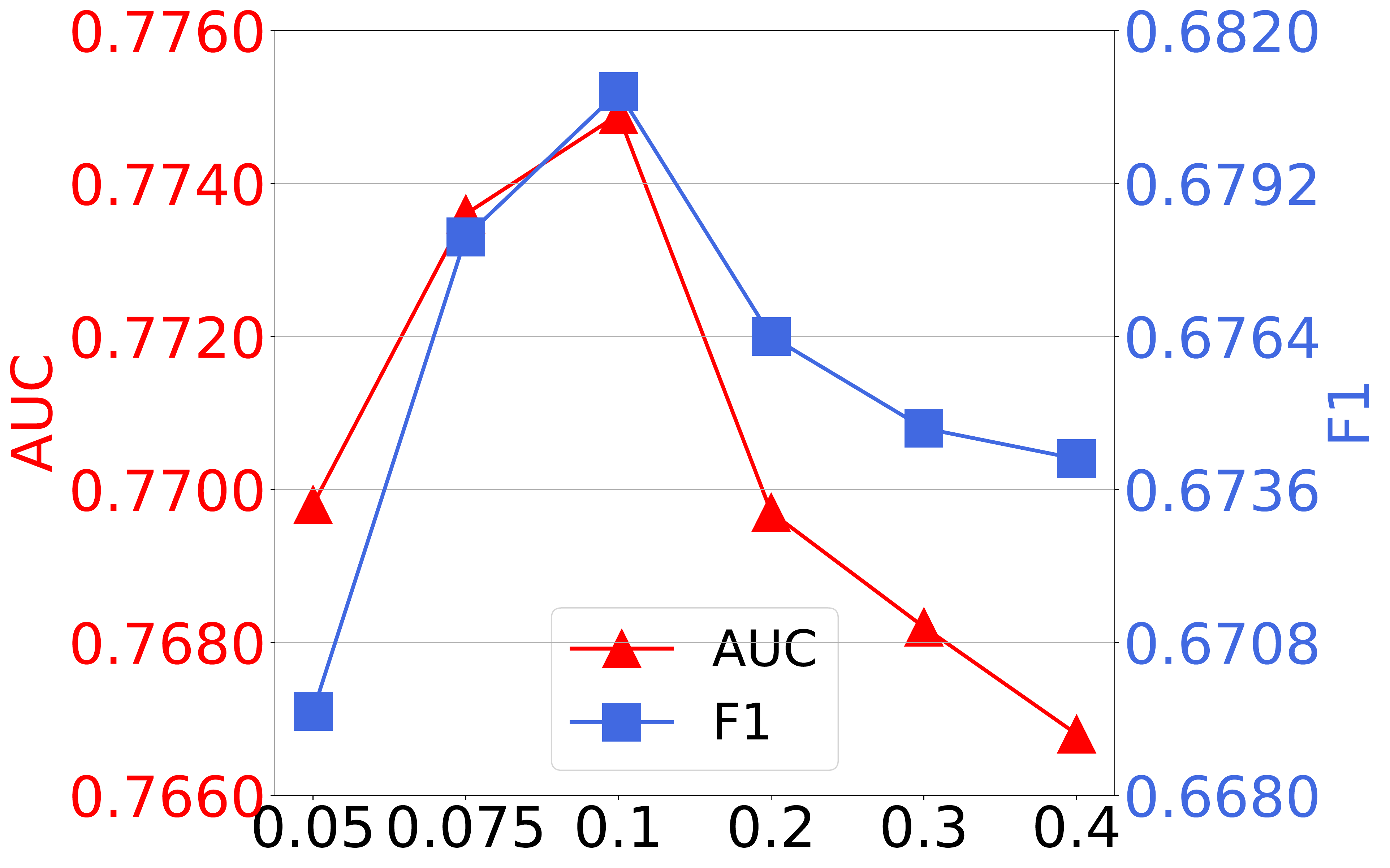} 
        \end{minipage}}
    \subfloat[Music]
    {   \begin{minipage}[t]{0.5\linewidth}
            \centering      
            \includegraphics[width=0.8\textwidth]{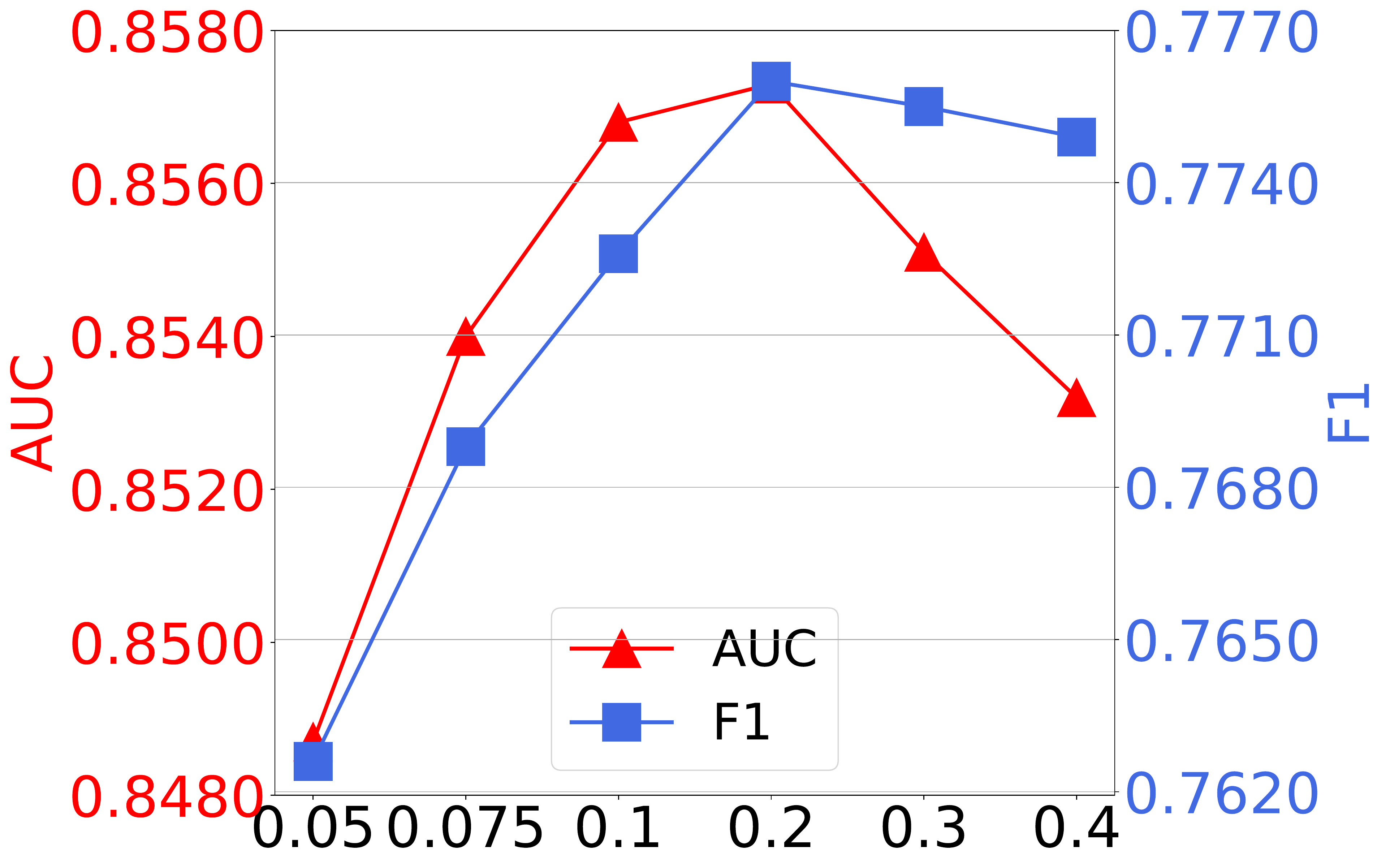}
        \end{minipage}}%
    \vspace{-0.2cm}
    \caption{Impact of temperature $\tau$ }
    \label{fig:tau} 
\end{figure}

In the overall contrastive loss defined in Equation~\eqref{overall_loss}, the trade-off parameter $\alpha$ can balance the influence of intra- and inter-graph contrastive losses. To analyze the influence of coefficient $\alpha$, we vary $\alpha$ in \{0.1, 0.5, 1, 1.5, 2,\} and report the results in Figure~\ref{fig:alpha}. From the results we could observe that: an appropriate $\alpha$ can effectively improve the performance of contrastive learning. Specifically, the model performs best when $\alpha = 1$, which means there exists equal importance between the intra- and inter-graph contrastive learning. In addition, with different $\alpha$, the performance of our KGIC is consistently better than other baselines, which also confirms the effectiveness of our multi-level interactive contrastive learning mechanism.

\subsubsection{Impact of Temperature $\tau$.}
As mentioned in previous contrastive learning work \cite{zhu2020deep, wu2021self}, the temperature $\tau$ defined in Equation~\eqref{intra_loss} and Equation~\eqref{inter_loss} plays an important role in contrastive learning. To investigate the impact of $\tau$, we vary it in range of \{0.05, 0.075, 0.1, 0.2, 0.3, 0.4\}. From the results shown in Figure~\ref{fig:tau}, we can find that: a too large value of $\tau$ will cause poor performance, consistent with conclusions of previous works \cite{wu2021self}. And generally, a temperature in the range of [0.1, 0.2] could lead to a satisfactory recommendation performance.

\subsection{Visualization (RQ4)}

\begin{figure}[tb] 
    \centering  
    \subfloat[Book]
    {   \begin{minipage}[t]{0.25\linewidth}
            \centering          
            \includegraphics[width=\textwidth]{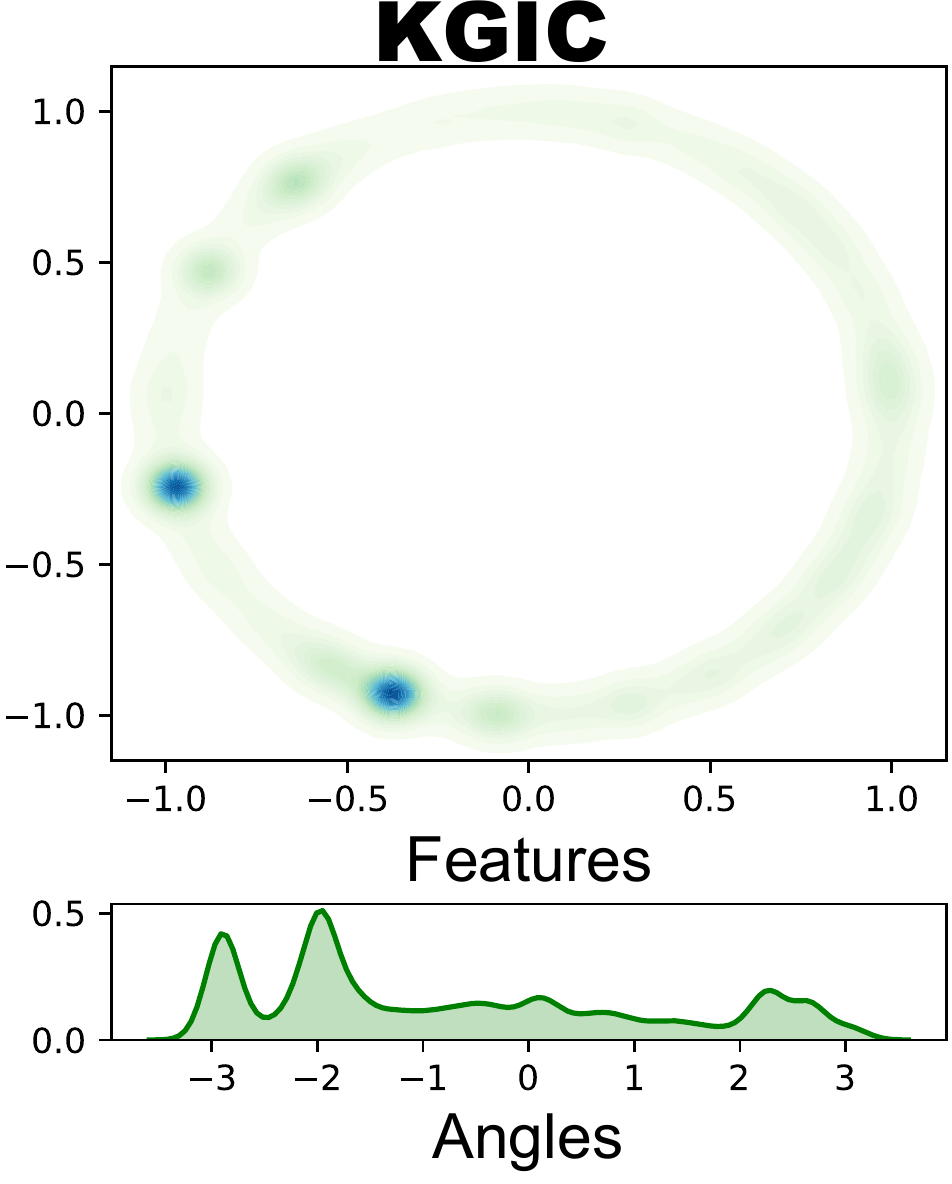}
        \end{minipage}
        \begin{minipage}[t]{0.25\linewidth}
            \centering          
            \includegraphics[width=\textwidth]{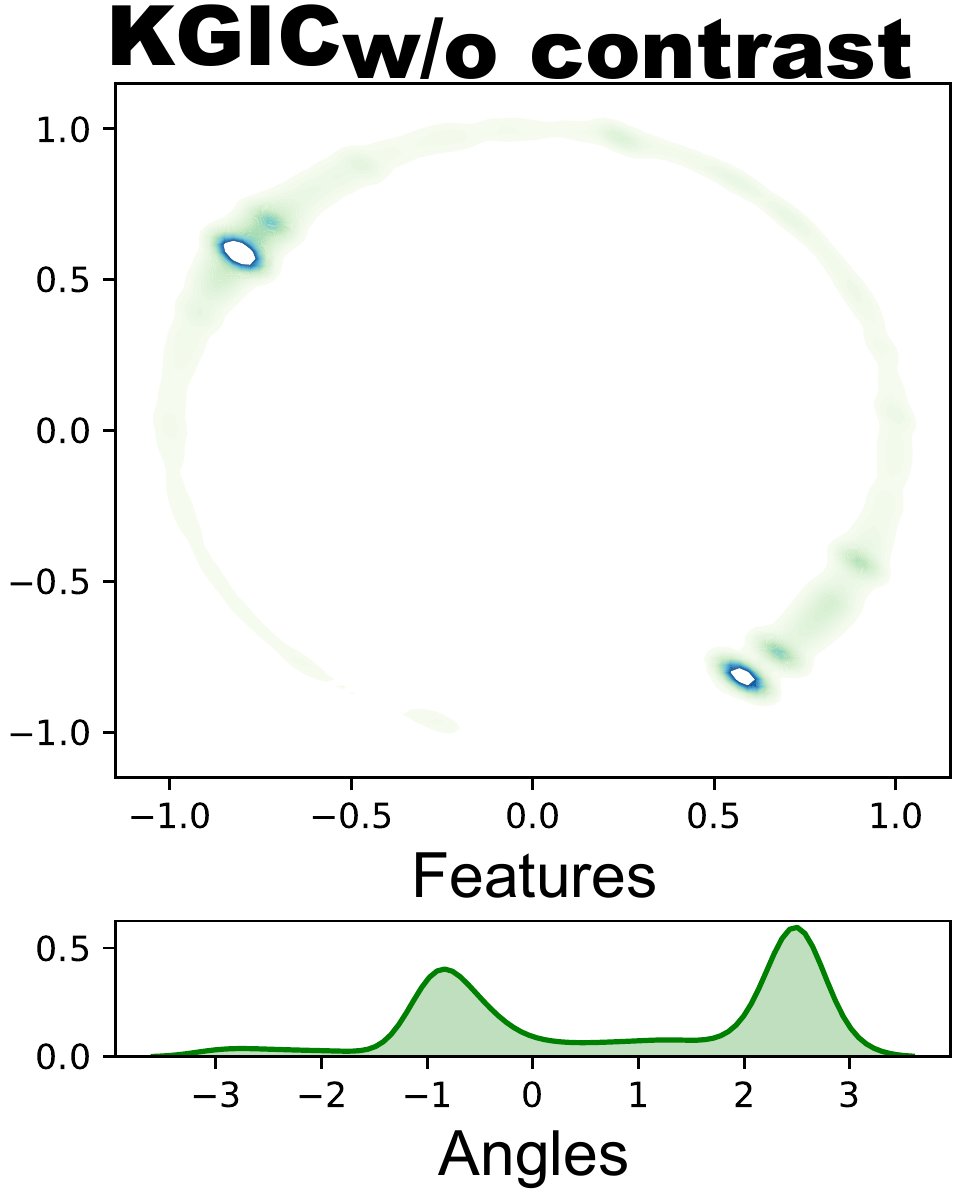}
        \end{minipage}
        \begin{minipage}[t]{0.25\linewidth}
            \centering          
            \includegraphics[width=\textwidth]{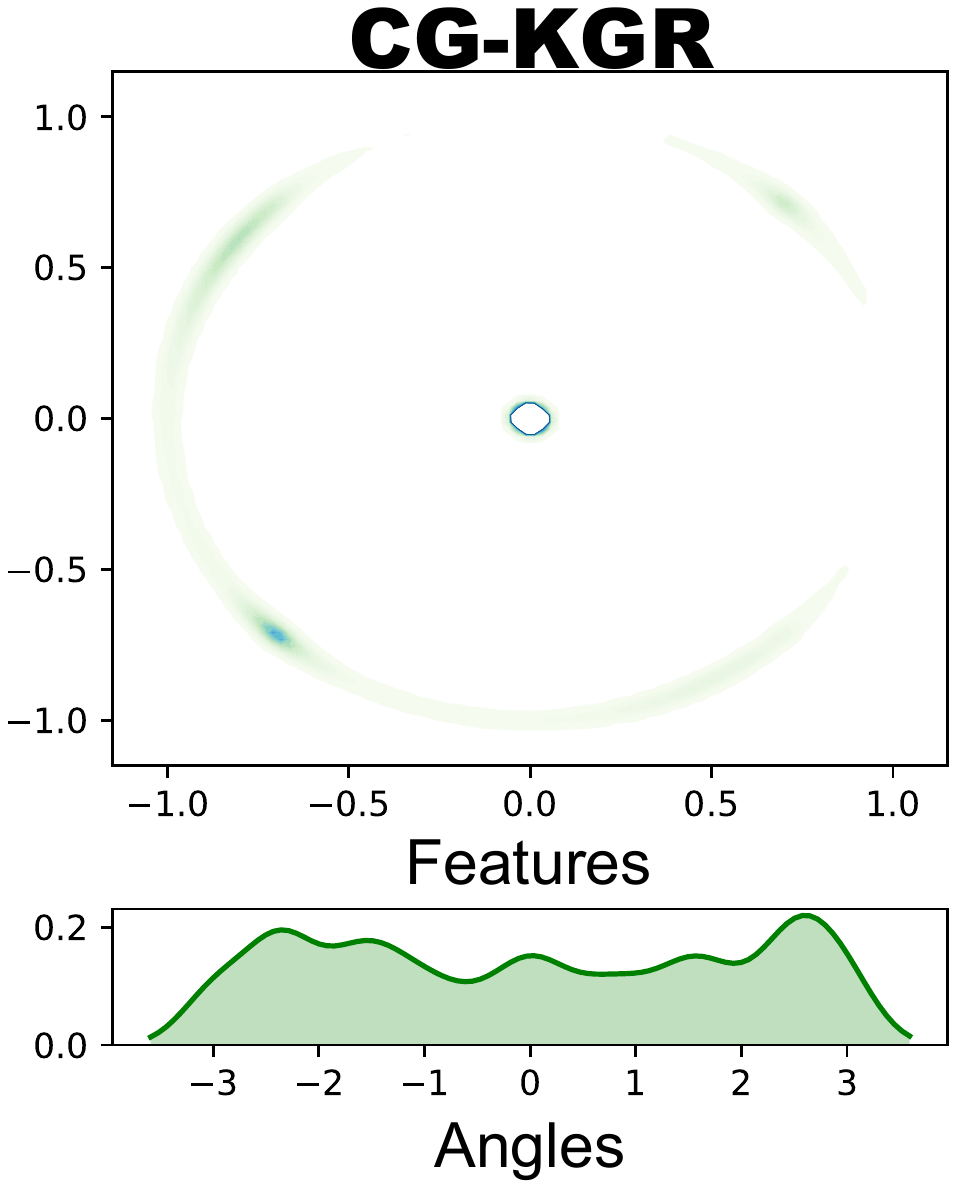}
        \end{minipage}
        \begin{minipage}[t]{0.25\linewidth}
            \centering          
            \includegraphics[width=\textwidth]{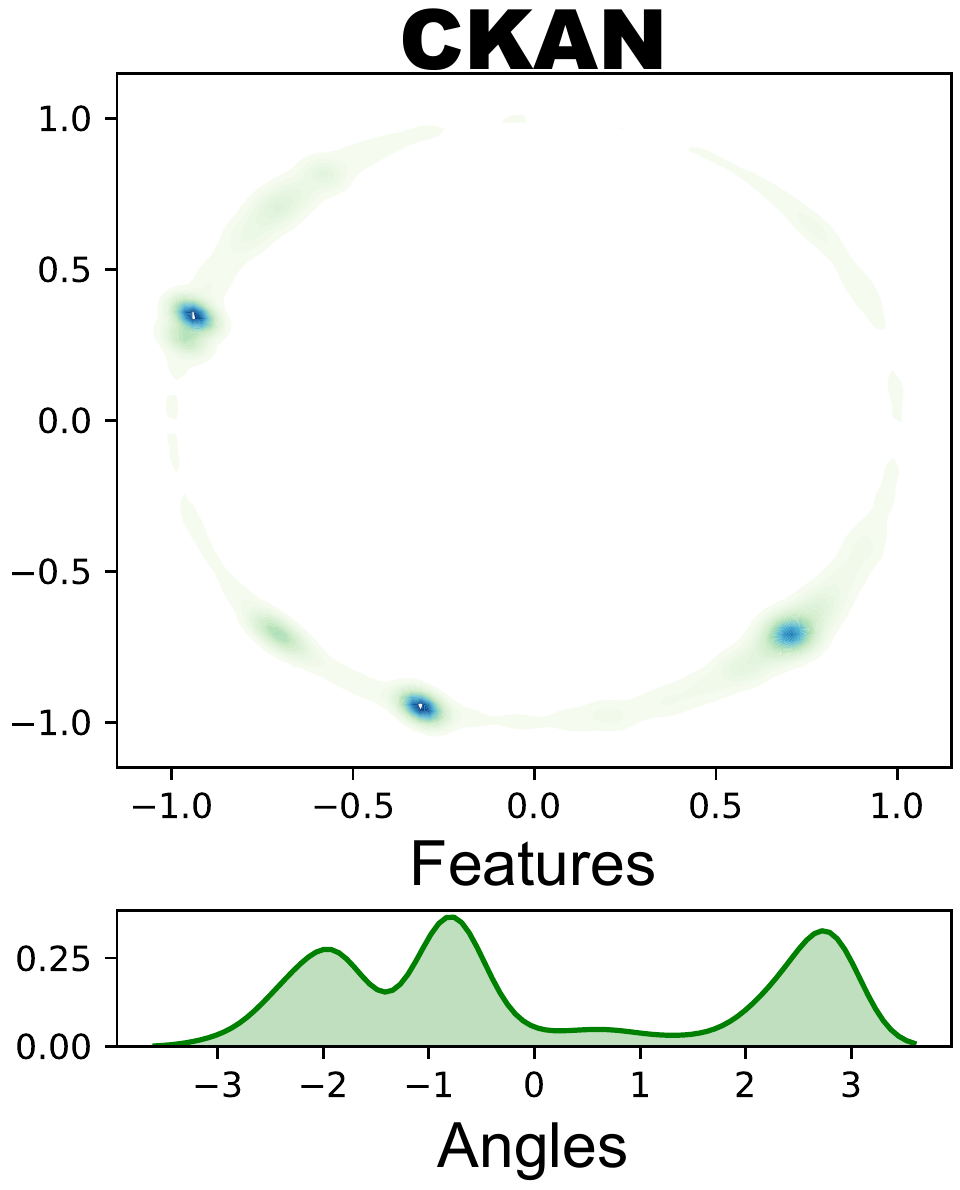}
        \end{minipage}
        }\\
    \vspace{-0.2cm}
    \subfloat[Music]
    {   \begin{minipage}[t]{0.25\linewidth}
            \centering      
            \includegraphics[width=\textwidth]{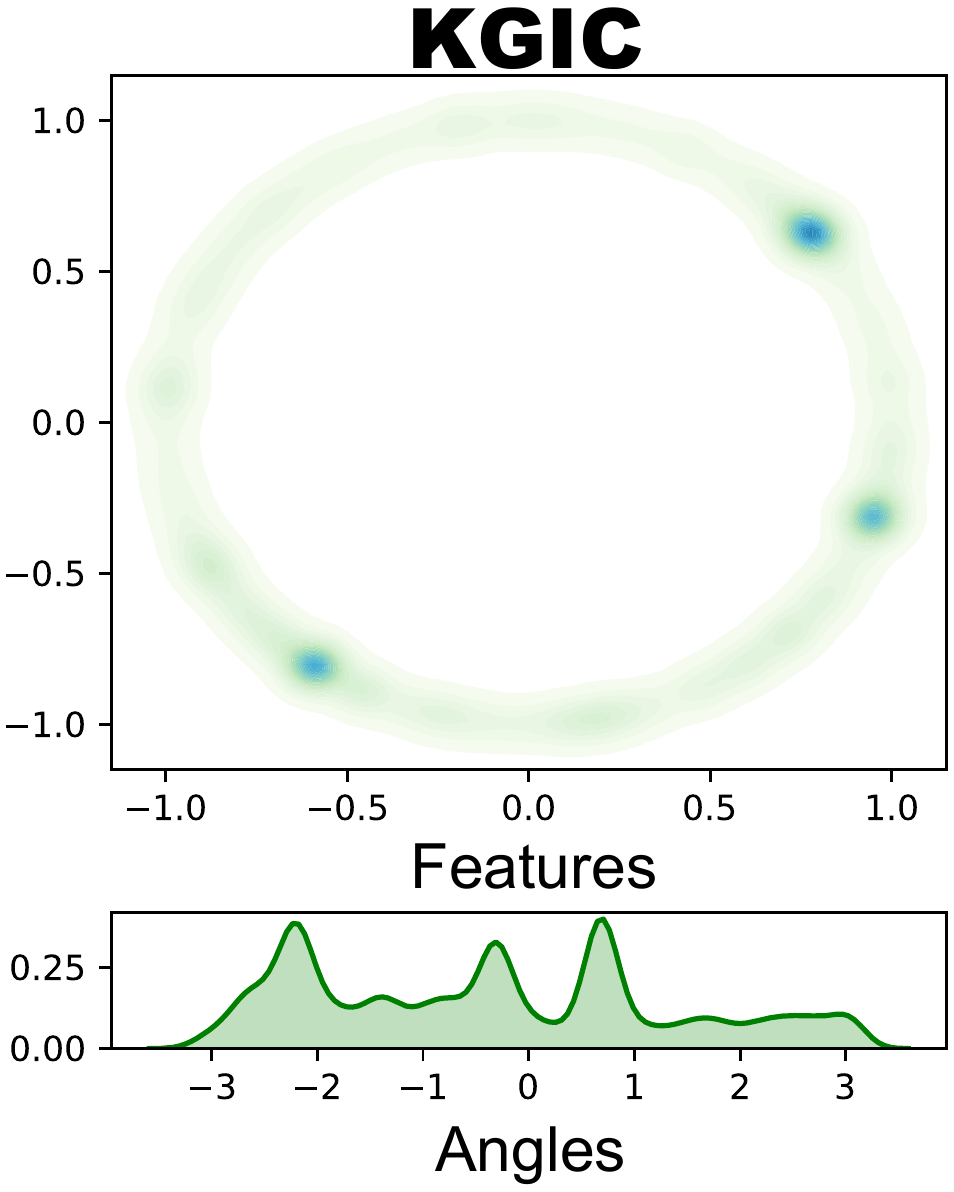}
        \end{minipage}
        \begin{minipage}[t]{0.25\linewidth}
            \centering      
            \includegraphics[width=\textwidth]{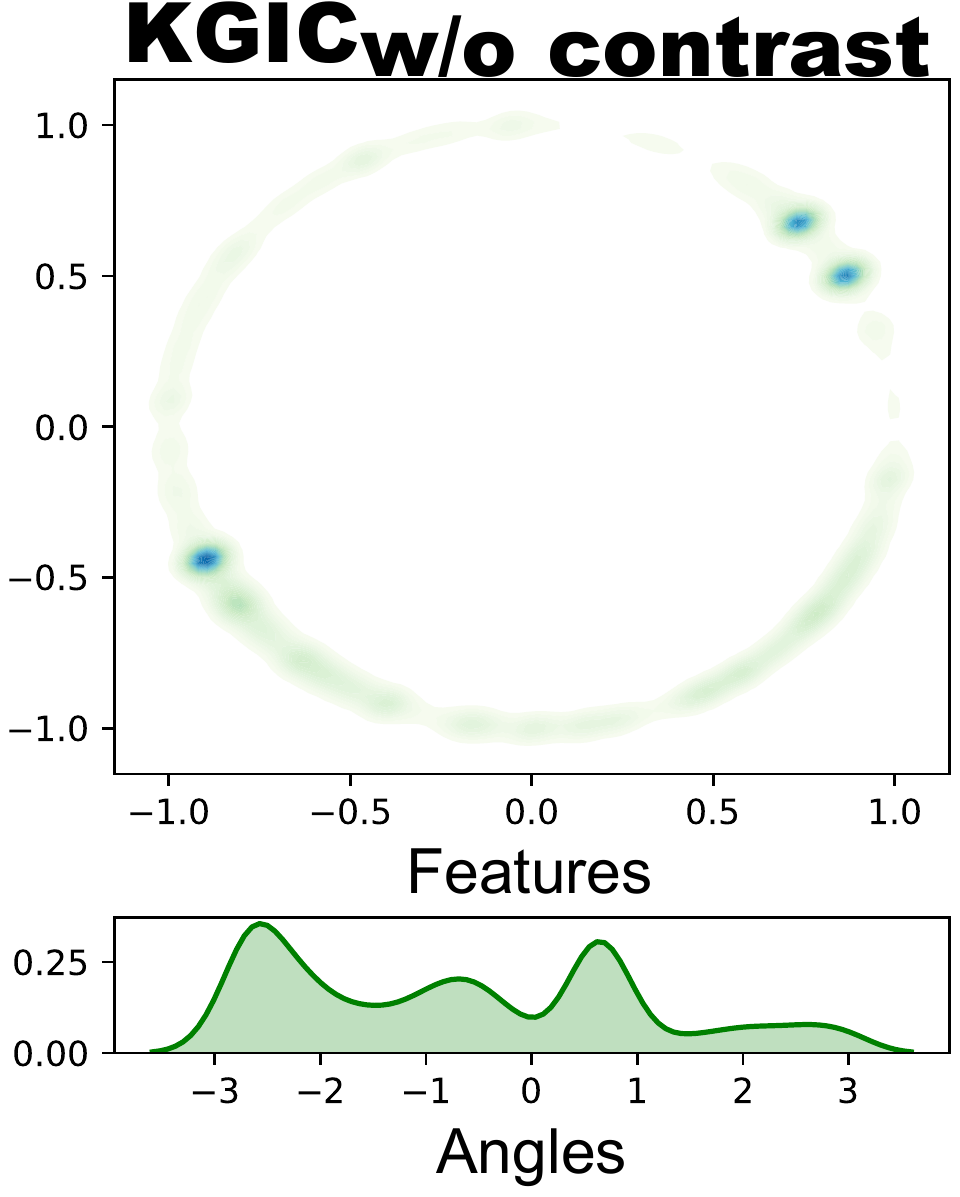}
        \end{minipage}
        \begin{minipage}[t]{0.25\linewidth}
            \centering      
            \includegraphics[width=\textwidth]{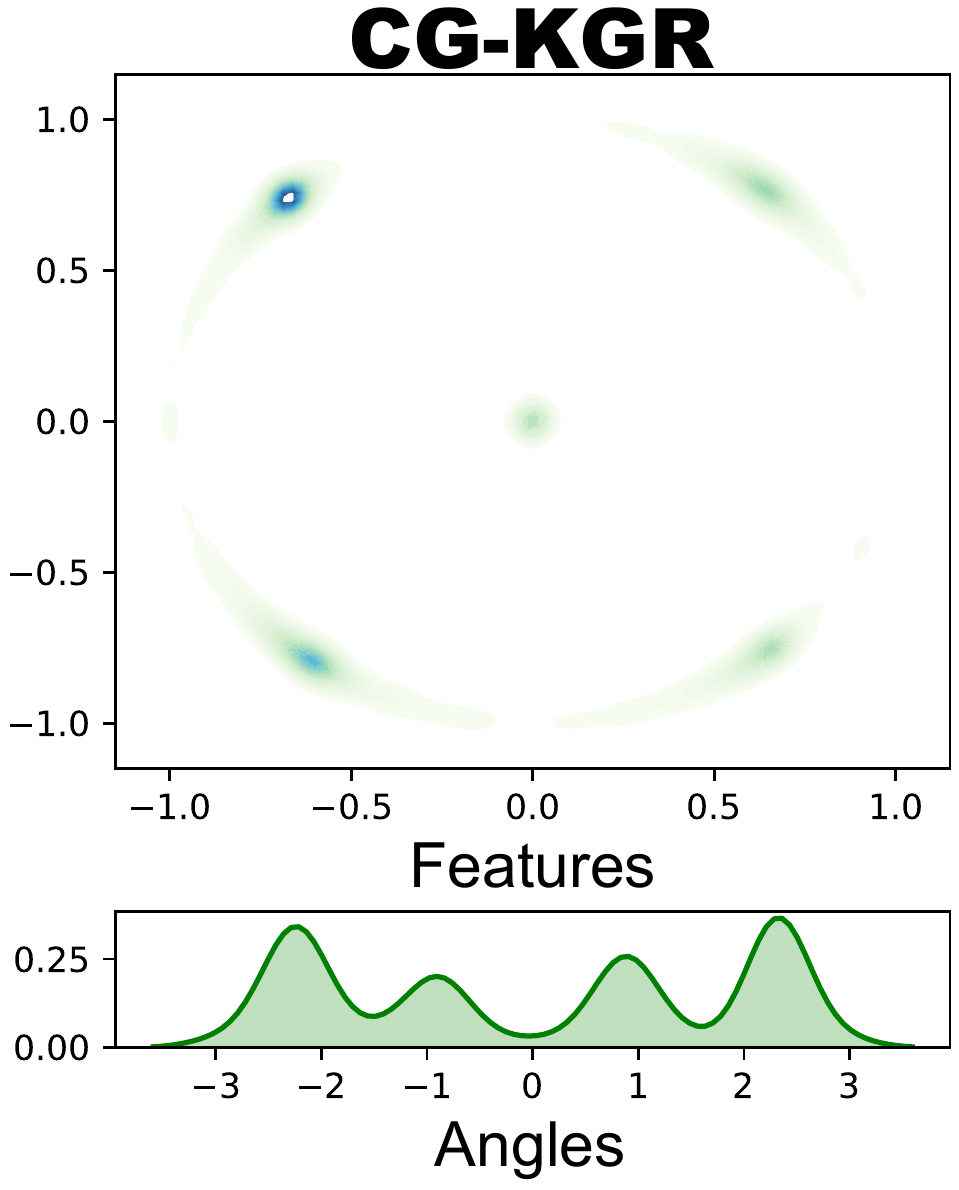}
        \end{minipage}
        \begin{minipage}[t]{0.25\linewidth}
            \centering      
            \includegraphics[width=\textwidth]{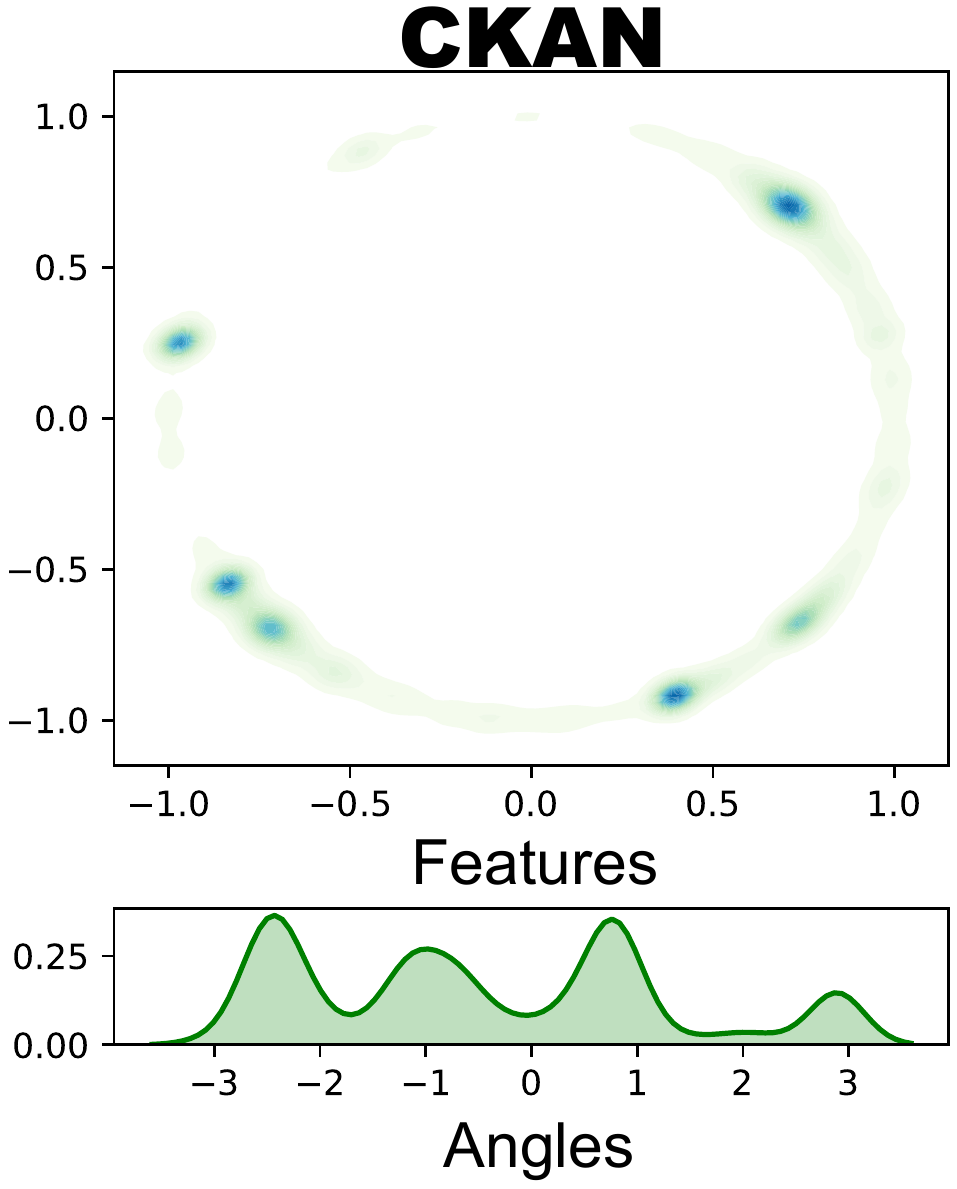}
        \end{minipage}
        }
    \vspace{-0.2cm}
    \caption{Visualization for the distribution of item embeddings.}
    \label{fig:visual} 
\end{figure}

Towards a more intuitive evaluation on how the proposed interactive contrastive mechanism affects the representation learning performance, we visualize the learned item embeddings in Figure~\ref{fig:visual}. Following previous Contrastive learning work \cite{yu2022are, lin2022ncl}, we plot item embeddings distributions with Gaussian kernel density estimation (KDE) \cite{botev2010kernel} in two-dimensional space (where the darker the color is, the more points fall in that area), and we also plot KDE on angles (\ie $arctan2(y, x)$ for each point $(x,y)$ in the above graph) towards a clearer presentation. 
As shown in Figure \ref{fig:visual}, we compare the visualized results of KGIC, $\text{KGIC}_{w/o \ contrast}$, CG-KGR, and CKAN on book and music, from which we have the following observations:
\begin{itemize}[leftmargin=*]
    \item CG-KGR and CKAN show highly clustered features that mainly reside on some narrow arcs, while our KGIC clearly has a more uniform distribution and hence is able to present more different node feature information. As mentioned in previous work \cite{wang2020understanding}, in a proper range, a more uniform distribution means a better capacity to model the diversity of node features.
    This fact demonstrates the superiority of KGIC in better representation learning and alleviating the representation degeneration problem.
    \item By removing all of the contrastive loss in KGIC, the learned item embeddings are less uniform and fall into several coherent clusters. This phenomenon confirms the promotion of representation learning comes from the proposed interactive contrastive mechanism, for preserving maximal information on representations and improving the uniformity of the learned representations.
\end{itemize}

\section{Conclusion}
In this paper, we focus on incorporating contrastive learning into KG-aware recommendation, making sufficient and coherent use of CF and KG in a self-supervised manner. We propose a novel framework, KGIC, which achieves better user/item representation learning from two dimensions:
(1) KGIC makes coherent utilization of CF and KG information in each local/non-local graph, by performing intra-graph interactive contrastive learning which contrasts layers of the CF and KG parts.
(2) KGIC sufficiently extracts and integrates more KG facts for user/item representation learning, by constructing non-local graphs and performing inter-graph interactive contrastive learning which contrasts the local and non-local graphs.
Extensive experiments on three public datasets demonstrate that KGIC significantly improves the recommendation performance over baselines on both Click-Through rate prediction and Top-K recommendation tasks.

\begin{acks}
    This work was supported in part by the National Natural Science Foundation of China under Grant No.61602197, Grant No.L1924068, Grant No.61772076, in part by CCF-AFSG Research Fund under Grant No.RF20210005, and in part by the fund of Joint Laboratory of HUST and Pingan Property \& Casualty Research (HPL). 
\end{acks}

\bibliographystyle{ACM-Reference-Format}
\balance
\bibliography{acmart}
\end{document}